\documentclass[prd,aps,amsmath,showpacs,superscriptaddress,twocolumn,endfloat]{revtex4-1}
\usepackage{graphicx}
\usepackage{color}
\usepackage{float}
\usepackage{subcaption}

\begin{document}

\title{Precision measurement of the $\eta \to \pi^+ \pi^- \pi^0$
   Dalitz plot distribution \\ with the KLOE detector}

\newcommand{\affinfnuniA}[3]{Dipartimento di Fisica dell'Universit\`a #1\\and INFN Sezione di #2, #3, Italy}
\newcommand{\affinfnuniB}[3]{Dipartimento di Scienze Fisiche dell'Universit\`a #1\\and INFN Sezione di #2, #3, Italy}
\newcommand{\affinfnuniC}[3]{Dipartimento di Fisica dell'Universit\`a #1\\and INFN gruppo collegato di #2, #3, Italy}
\newcommand{\affuni}[2]{Dipartimento di Fisica dell'Universit\`a #1, #2, Italy.}
\newcommand{\affinfn}[2]{INFN Sezione di #1, #2, Italy.}
\newcommand*{\INFNBari}{\affinfn{Bari}{Bari}}
\newcommand*{\INFNCatania}{\affinfn{Catania}{Catania}}
\newcommand*{\Cracow}{Institute of Physics, Jagiellonian University, Cracow, Poland.}
\newcommand*{\Frascati}{Laboratori Nazionali di Frascati dell'INFN, Frascati, Italy.}
\newcommand*{\Messina}{Dipartimento di Fisica e Scienze della Terra dell'Universit\`a di Messina, Messina, Italy.}
\newcommand*{\MessinaII}{Dipartimento di Scienze Chimiche, Biologiche, Farmaceutiche ed Ambientali dell'Universit\`a di Messina, Messina, Italy.}
\newcommand*{\INFNMessina}{INFN Gruppo collegato di Messina, Messina, Italy.}
\newcommand*{\Calabria}{\affuni{della Calabria}{Rende}}
\newcommand*{\INFNCalabria}{INFN Gruppo collegato di Cosenza, Rende, Italy.}
\newcommand*{\Energetica}{Dipartimento di Scienze di Base ed Applicate per l'Ingegneria dell'Universit\`a 
``Sapienza'', Roma, Italy.}
\newcommand*{\Marconi}{Dipartimento di Scienze e Tecnologie applicate, Universit\`a ``Guglielmo Marconi", Roma, Italy.}
\newcommand*{\Novosibirsk}{Novosibirsk State University, 630090 Novosibirsk, Russia.}
\newcommand*{\RomaI}{\affuni{``Sapienza''}{Roma}}
\newcommand*{\INFNRomaI}{\affinfn{Roma}{Roma}}
\newcommand*{\RomaII}{\affuni{``Tor Vergata''}{Roma}}
\newcommand*{\INFNRomaII}{\affinfn{Roma Tor Vergata}{Roma}}
\newcommand*{\RomaIII}{Dipartimento di Matematica e Fisica dell'Universit\`a 
``Roma Tre'', Roma, Italy.}
\newcommand*{\INFNRomaIII}{\affinfn{Roma Tre}{Roma}}
\newcommand*{\ENEACasaccia}{ENEA UTTMAT-IRR, Casaccia R.C., Roma, Italy}
\newcommand*{\Uppsala}{Department of Physics and Astronomy, Uppsala University, Uppsala, Sweden.}
\newcommand*{\Warsaw}{National Centre for Nuclear Research, Warsaw, Poland.}

\collaboration{The KLOE-2 Collaboration}\noaffiliation
\author{A.~Anastasi}\affiliation{\Messina}\affiliation{\Frascati}
\author{D.~Babusci}\affiliation{\Frascati}
\author{G.~Bencivenni}\affiliation{\Frascati}
\author{M.~Berlowski}\affiliation{\Warsaw}
\author{C.~Bloise}\affiliation{\Frascati}
\author{F.~Bossi}\affiliation{\Frascati}
\author{P.~Branchini}\affiliation{\INFNRomaIII}
\author{A.~Budano}\affiliation{\RomaIII}\affiliation{\INFNRomaIII}
\author{L.~Caldeira~Balkest\aa hl}\affiliation{\Uppsala}
\author{B.~Cao}\affiliation{\Uppsala}
\author{F.~Ceradini}\affiliation{\RomaIII}\affiliation{\INFNRomaIII}
\author{P.~Ciambrone}\affiliation{\Frascati}
\author{F.~Curciarello}\affiliation{\Messina}\affiliation{\INFNCatania}\affiliation{\Novosibirsk}
\author{E.~Czerwi\'nski}\affiliation{\Cracow}
\author{G.~D'Agostini}\affiliation{\RomaI}\affiliation{\INFNRomaI}
\author{E.~Dan\`e}\affiliation{\Frascati}
\author{V.~De~Leo}\affiliation{\INFNRomaIII}
\author{E.~De~Lucia}\affiliation{\Frascati}
\author{A.~De~Santis}\affiliation{\Frascati}
\author{P.~De~Simone}\affiliation{\Frascati}
\author{A.~Di~Cicco}\affiliation{\RomaIII}\affiliation{\INFNRomaIII}
\author{A.~Di~Domenico}\affiliation{\RomaI}\affiliation{\INFNRomaI}
\author{R.~Di~Salvo}\affiliation{\INFNRomaII}
\author{D.~Domenici}\affiliation{\Frascati}
\author{A.~D'Uffizi}\affiliation{\Frascati}
\author{A.~Fantini}\affiliation{\RomaII}\affiliation{\INFNRomaII}
\author{G.~Felici}\affiliation{\Frascati}
\author{S.~Fiore}\affiliation{\ENEACasaccia}\affiliation{\INFNRomaI}
\author{A.~Gajos}\affiliation{\Cracow}
\author{P.~Gauzzi}\affiliation{\RomaI}\affiliation{\INFNRomaI}
\author{G.~Giardina}\affiliation{\Messina}\affiliation{\INFNCatania}
\author{S.~Giovannella}\affiliation{\Frascati}
\author{E.~Graziani}\affiliation{\INFNRomaIII}
\author{F.~Happacher}\affiliation{\Frascati}
\author{L.~Heijkenskj\"old}\affiliation{\Uppsala}
\author{W.~Ikegami Andersson}\affiliation{\Uppsala}
\author{T.~Johansson}\affiliation{\Uppsala}
\author{D.~Kami\'nska}\affiliation{\Cracow}
\author{W.~Krzemien}\affiliation{\Warsaw}
\author{A.~Kupsc}\affiliation{\Uppsala}
\author{S.~Loffredo}\affiliation{\RomaIII}\affiliation{\INFNRomaIII}
\author{G.~Mandaglio}\affiliation{\MessinaII}\affiliation{\INFNMessina}
\author{M.~Martini}\affiliation{\Frascati}\affiliation{\Marconi}
\author{M.~Mascolo}\affiliation{\Frascati}
\author{R.~Messi}\affiliation{\RomaII}\affiliation{\INFNRomaII}
\author{S.~Miscetti}\affiliation{\Frascati}
\author{G.~Morello}\affiliation{\Frascati}
\author{D.~Moricciani}\affiliation{\INFNRomaII}
\author{P.~Moskal}\affiliation{\Cracow}
\author{M.~Papenbrock}\affiliation{\Uppsala}
\author{A.~Passeri}\affiliation{\INFNRomaIII}
\author{V.~Patera}\affiliation{\Energetica}\affiliation{\INFNRomaI}
\author{E.~Perez~del~Rio}\affiliation{\Frascati}
\author{A.~Ranieri}\affiliation{\INFNBari}
\author{P.~Santangelo}\affiliation{\Frascati}
\author{I.~Sarra}\affiliation{\Frascati}
\author{M.~Schioppa}\affiliation{\Calabria}\affiliation{\INFNCalabria}
\author{M.~Silarski}\affiliation{\Frascati}
\author{F.~Sirghi}\affiliation{\Frascati}
\author{L.~Tortora}\affiliation{\INFNRomaIII}
\author{G.~Venanzoni}\affiliation{\Frascati}
\author{W.~Wi\'slicki}\affiliation{\Warsaw}
\author{M.~Wolke}\affiliation{\Uppsala}



\begin{abstract}
Using $1.6$ fb$^{-1}$ of $e^+ e^-\to\phi\to\eta\gamma$ data collected
with the KLOE detector at DA$\Phi$NE, the Dalitz plot distribution for
the $\eta \to \pi^+ \pi^- \pi^0$ decay is studied with the world's
largest sample of $\sim 4.7 \cdot 10^6$ events.  The Dalitz plot
density is parametrized as a polynomial expansion up to cubic terms in
the normalized dimensionless variables $X$ and $Y$. The experiment is
sensitive to all charge conjugation conserving terms of the expansion,
including a $gX^2Y$ term. The statistical uncertainty of all
parameters is improved by a factor two with respect to earlier
measurements.

\end{abstract}


\maketitle

\section{Introduction}
The isospin violating $\eta \to \pi^+ \pi^- \pi^0$ decay can proceed
via electromagnetic interactions or via strong interactions due to
the difference between the masses of $u$ and $d$ quarks. The
electromagnetic part of the decay amplitude is long known to be
strongly suppressed \cite{sutherland66,sutherland_bell68}. The recent
calculations performed at next-to-leading order (NLO) of the chiral
perturbation theory (ChPT) \cite{baur_kambor_wyler96,
  ditsche_kubis_meissner2009} reaffirm that the decay amplitude is
dominated by the isospin violating part of the strong interaction.

Defining the quark mass ratio, $Q$, as 
\begin{equation}
 Q^2 \equiv \frac{m_s^2 - \hat{m}^2}{m_d^2 - m_u^2} \qquad \text{ with }\hat{m}= \frac{1}{2}(m_d+m_u), \label{eq:q}
\end{equation}
the decay width at up to NLO ChPT is proportional to $Q^{-4}$
\cite{bijnens_gasser2002}. The definition in Eq.~(\ref{eq:q}),
neglecting $\hat{m}^2/m_s^2$, gives an ellipse in the $m_s/m_d,
m_u/m_d$ plane with major semi-axis $Q$ \cite{leutwyler96}: a
determination of $Q$ puts a stringent constraint on the light quark
masses.

Using Dashen's theorem \cite{dashen69} to account for the
electromagnetic effects, $Q$ can be determined at the lowest order from
a combination of kaon and pion masses. With this value of $Q=24.2$,
the ChPT results for the $\eta \to \pi^+ \pi^- \pi^0$ decay width
at LO, $\Gamma_{LO}=66$ eV, and NLO, $\Gamma_{NLO}=160\pm50$~eV
\cite{gasser_leutwyler85_eta3pi}, are far from the experimental value
$\Gamma_{exp}=300\pm11$~eV~\cite{PDG14}.  The discrepancy could
originate from higher order contributions to the decay amplitude or
from the corrections to the $Q$ value.

Several theoretical efforts have aimed at a better description of the
$\eta \to 3\pi$ decay amplitude.  A full NNLO ChPT calculation has
been performed \cite{bijnens_ghorbani2007}. However, it depends on the
values of a large number of the coupling constants of the chiral lagrangian 
which are not known precisely and 
therefore is affected by large uncertainties.  A calculation in unitarized ChPT
using relativistic coupled channels \cite{borasoy_nissler2005} gives
better agreement with the data but it is model dependent. With a
non-relativistic effective field theory framework, the effects of
higher order isospin breaking in the final state interactions have
been investigated \cite{schneider_kubis_ditsche2011}. The $\pi \pi$
rescattering seems to play an important role in this decay, giving
about half of the correction in going from LO to NLO
\cite{gasser_leutwyler85_eta3pi}.  The rescattering can be accounted for to
all orders using dispersive integrals. In this framework, there are
three approaches to improve the ChPT predictions
\cite{kampf_knecht_novotny_zdrahal2011,colangelo_lanz_leutwyler_passemar2011,guo2015}.
The reliability of the ChPT calculations could be checked by a
comparison with the experimental pion distributions represented by the
Dalitz plot. Conversely, precise experimental distributions could be used 
directly for
the dispersive analyses
\cite{kampf_knecht_novotny_zdrahal2011,colangelo_lanz_leutwyler_passemar2011,guo2015} to 
determine the $Q$ ratio without relying on the higher order ChPT
calculations.  

For the $\eta \to \pi^+ \pi^- \pi^0$ Dalitz plot distribution, the normalized
variables $X$ and $Y$ are commonly used:
 \begin{align}
X &= \sqrt{3} \frac{T_{\pi^+} - T_{\pi^-}}{Q_\eta}\label{eq:XasT}\\
Y &= \frac{3T_{\pi^0}}{Q_\eta} -1 \label{eq:YasT} 
\end{align}
with
\begin{align}
Q_\eta & =T_{\pi^+} +  T_{\pi^-} + T_{\pi^0} = m_\eta - 2m_{\pi^+} -m_{\pi^0} . \label{eq:Q}
\end{align}
$T_i$ are kinetic energies of the pions in the $\eta$ rest frame.
The  squared amplitude of the decay is parametrized by a polynomial expansion around $(X,Y)=(0,0)$:
\begin{eqnarray}\label{eq:DPamplitude}
|A(X,Y)|^2&\simeq& N(1+\\ \nonumber
&&aY+bY^2+cX+dX^2+eXY + \\ \nonumber
&&fY^3+ gX^2Y + hX Y^2 + lX^3 + \ldots).
\end{eqnarray}
The Dalitz plot distribution can then be fit using this formula to
extract the parameters $a,b,\ldots$, usually called the Dalitz plot
parameters.  Note that coefficients multiplying odd powers of $X$
($c,e,h$ and $l$) must be zero assuming charge conjugation invariance.

The experimental values of the Dalitz plot parameters are shown in
Tab.~\ref{tab:prev_results} together with the parametrization of
theoretical calculations. The most precise previous measurement is
from KLOE 2008 analysis which was based on $1.34\cdot 10^6$ events
\cite{kloe2008}.  There is some disagreement among the experiments,
specially for the $b$ but also for the $a$ parameter. Looking at the
theory, both the $b$ and the $f$ parameters deviate from
experiment. The new high statistics measurement presented in this
paper can help to clarify the tension among the experimental results,
and can be used as a more precise input for the dispersive
calculations.

\begin{table*}[tbhp]
\caption{Summary of Dalitz plot parameters from experiments and theoretical 
predictions.\label{tab:prev_results}}
\begin{tabular}{l | l l l l l}
\hline\hline
\textbf{Experiment} & \multicolumn{1}{c}{$-a$} &
\multicolumn{1}{c}{$ b$} & \multicolumn{1}{c}{$d$} & 
\multicolumn{1}{c}{$f$} & \multicolumn{1}{c}{$-g$} \\
\hline
Gormley(70)\hfill\cite{gormley70} & $1.17\pm0.02$ & $0.21\pm0.03$ & $0.06\pm0.04$ & $-$ & $-$\\
Layter(73)\hfill\cite{layter73} & $1.080\pm0.014$ & $0.03\pm0.03$ & $0.05\pm0.03$ & $-$ & $-$\\
CBarrel(98)\hfill\cite{CBarrel98} & $1.22\pm0.07$ & $0.22\pm0.11$ & $0.06$(fixed) & $-$&$-$\\
KLOE(08)\hfill\cite{kloe2008} & $1.090\pm0.005^{+0.019}_{-0.008}$ & $0.124\pm0.006\pm0.010$ & $0.057\pm0.006^{+0.007}_{-0.016}$ & $0.14\pm0.01\pm0.02$&$-$ \\
WASA(14)\hfill\cite{patrik2014} & $1.144\pm0.018$ & $0.219\pm0.019\pm0.047$  & 
$0.086\pm0.018\pm0.015$& $0.115\pm0.037$&$-$\\
BESIII(15)\hfill\cite{besIII2015} &  $1.128\pm0.015\pm0.008$ & 
$0.153\pm0.017\pm0.004$ &$0.085\pm0.016\pm0.009$ & $0.173\pm0.028\pm0.021$& $-$ \\
\hline \hline
\textbf{Calculations} &  & &  &  &  \\ \hline
ChPT LO\hfill\cite{bijnens_ghorbani2007} & 1.039 & 0.27 & 0 & 0& $-$ \\
ChPT NLO\hfill\cite{bijnens_ghorbani2007} & 1.371 & 0.452 & 0.053 & 0.027 & $-$\\
ChPT NNLO\hfill\cite{bijnens_ghorbani2007} & $1.271\pm0.075$ & 
$0.394\pm0.102$ & $0.055\pm0.057$ & $0.025\pm0.160$ & $-$\\
dispersive\hfill\cite{kambor_wiesendanger_wyler96} & 1.16 & 0.26 & 0.10 &$-$ & $-$ \\
simplified disp\hfill\cite{bijnens_gasser2002} & 1.21 & 0.33 & 0.04 &$-$ & $-$\\
NREFT\hfill\cite{schneider_kubis_ditsche2011} & $1.213\pm0.014$ & $0.308\pm0.023$ & $0.050\pm0.003$ & $0.083\pm0.019$ & $0.039\pm0.002$\\
UChPT\hfill\cite{borasoy_nissler2005} & $1.054\pm0.025$ & $0.185\pm0.015$ & $0.079\pm0.026$ & $0.064\pm0.012$ & $-$\\
\hline\hline
\end{tabular} 
\end{table*}

\section{The KLOE detector}

The KLOE detector at the DA$\Phi$NE $e^+e^-$ collider in Frascati
consists of a large cylindrical Drift chamber (DC) and an
electromagnetic calorimeter (EMC) in a 0.52 T axial magnetic
field. The DC \cite{KLOEDCNIM} is 4 m in diameter and 3.3 m long and
is operated with a helium - isobutane gas mixture (90\% -
10\%). Charged particles are reconstructed with a momentum resolution
of $\sigma(p_\perp)/p_\perp \simeq 0.4\%$.

The EMC \cite{KLOEEMCNIM} consists of alternating layers of lead and scintillating fibers covering 98\% of the solid angle. The lead-fiber layers are arranged in $\sim (4.4 \times 4.4) \text{ cm}^2$ cells, five in depth, and these are read out at both ends. Hits in cells close in time and space are grouped together in clusters. Cluster energy is obtained from the signal amplitude and has a resolution  of $\sigma(E)/E = 5.7\%/\sqrt{E (\text{GeV)}}$. Cluster time, $t_{\rm cluster}$, and position  are energy weighted averages, with time resolution 
$\sigma(t) = (57 \text{ ps})/\sqrt{E (\text{GeV)}} \oplus 100$ ps. The cluster
position along the fibers is obtained from time differences of the signals.

The KLOE trigger \cite{KLOEtrigger} uses both EMC and DC
information. The trigger conditions are chosen to minimize beam
background. In this analysis, events are selected with the calorimeter
trigger, requiring two energy deposits with $E>50$ MeV for the barrel
and $E>150$ MeV for the endcaps.

The analysis is performed using data collected at the $\phi$ meson
peak with the KLOE detector in 2004-2005, and corresponds to an
integrated luminosity of $\sim 1.6 \text{ fb}^{-1}$.  Due to
DA$\Phi$NE crossing angle $\phi$ mesons
have a small horizontal momentum, {\bf p}$_\phi$ of about 13 MeV/c.
 The $\eta$ mesons are produced
in the radiative decay $\phi\to\eta\gamma_\phi $.  
The photon from the $\phi$ radiative decay, $\gamma_\phi$, has an energy
$E \sim 363$ MeV. The data sample used for this analysis
is independent and about four times larger than the one used in the
previous KLOE(08) $\eta \to \pi^+ \pi^- \pi^0$ Dalitz plot analysis \cite{kloe2008}.

The reconstructed data are sorted by an event classification procedure
which rejects beam and cosmic ray backgrounds and splits the events
into separate streams according to their
topology~\cite{KLOEdatahandling}. The beam and background conditions
are monitored. The corresponding parameters are stored for each run
and included in the \textsc{GEANT3} based Monte Carlo (MC) simulation
of the detector. The event generators for the production and decays of
the $\phi$-meson include simulation of initial state radiation. The
final state radiation is included for the simulation of the signal
process. The simulation of $e^+ e^- \to \omega \pi^0$ process (an
important background in this analysis) assumes a cross section of $8$
nb.  The simulations of the background channels used in this analysis
correspond to the integrated luminosity of the experimental data set,
while the signal simulation corresponds to ten times larger
luminosity.

\section{Event Selection}

\begin{figure*}[htbp]
        \begin{subfigure}[b]{0.5\textwidth}
                \includegraphics[width=\textwidth]{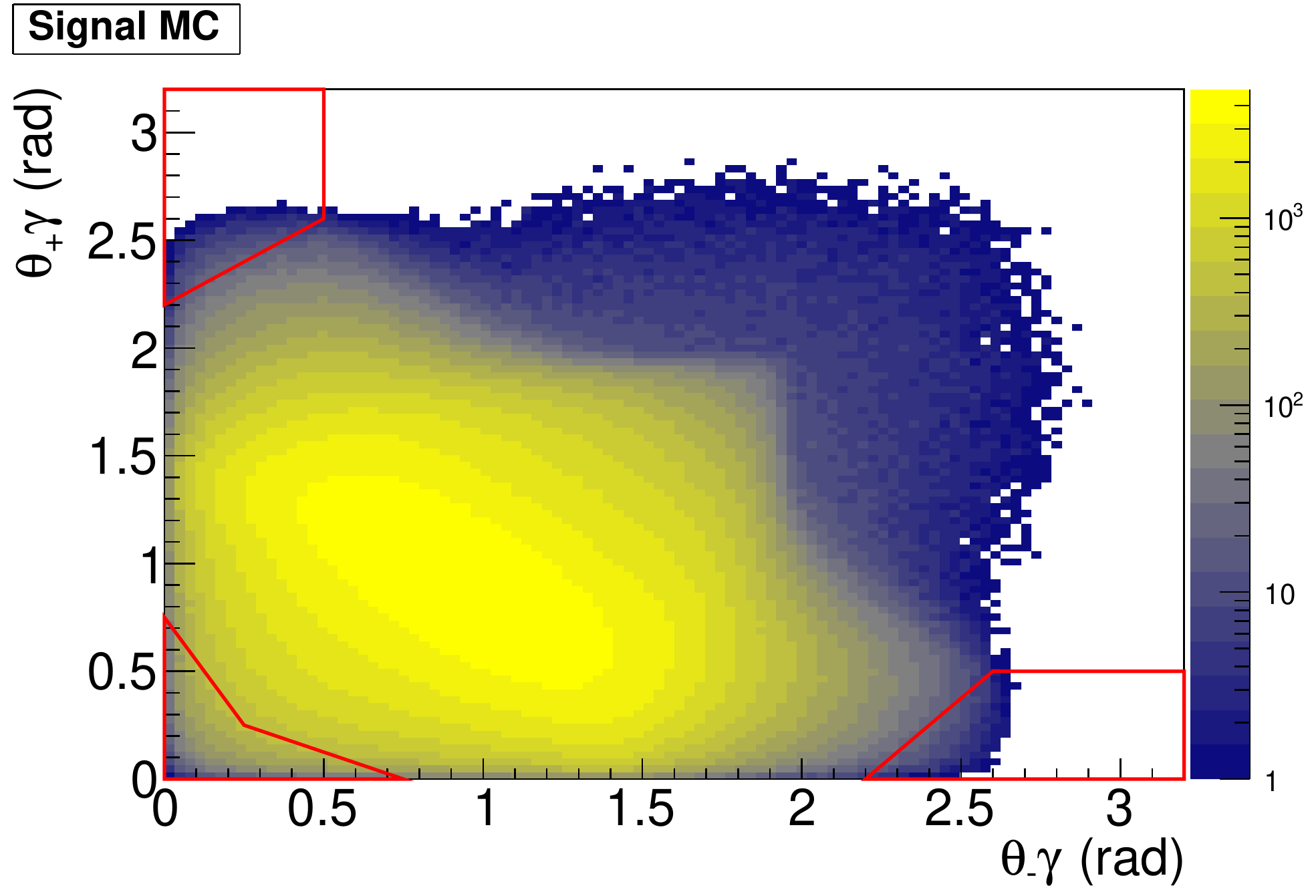}
        \end{subfigure}%
        ~ 
        \begin{subfigure}[b]{0.5\textwidth}
                \includegraphics[width=\textwidth]{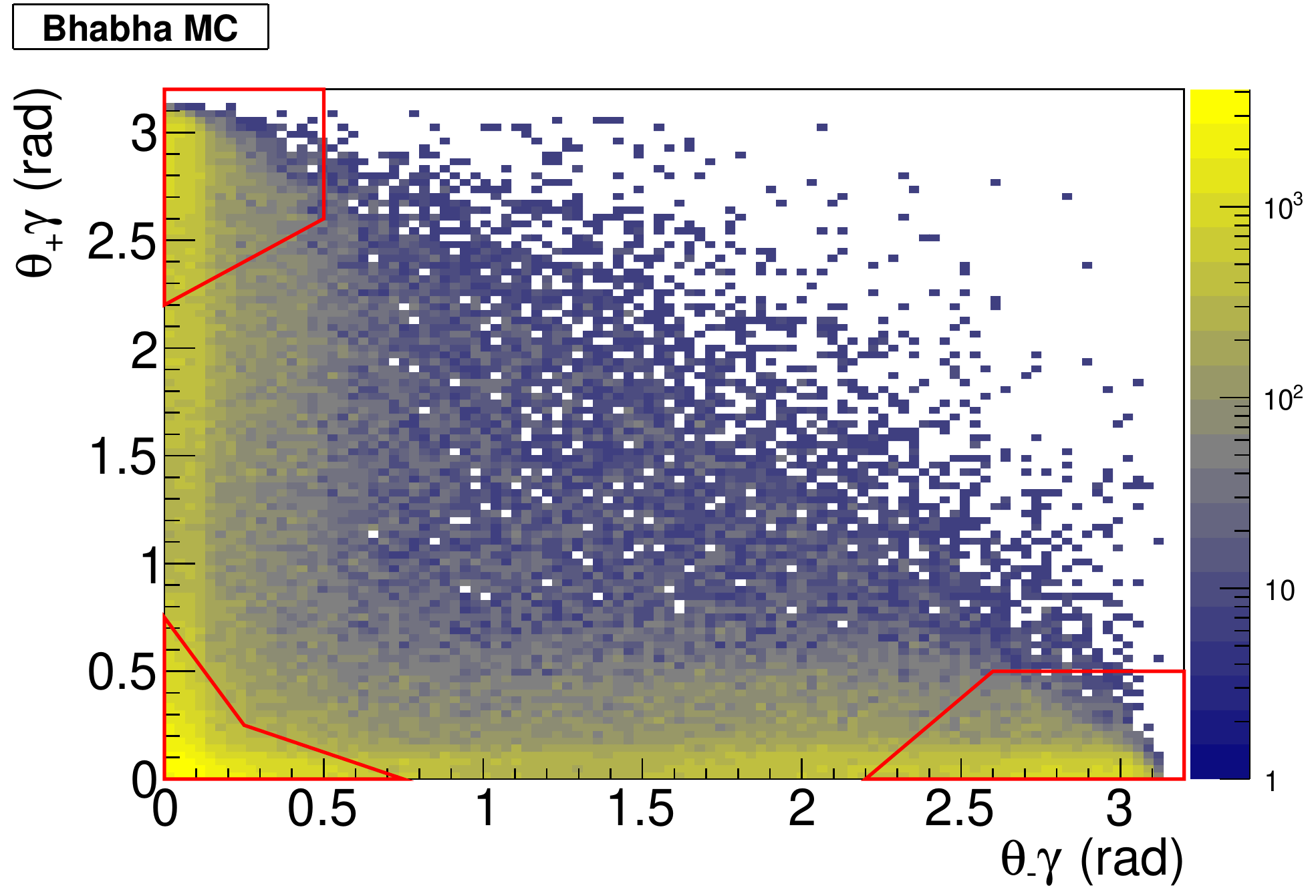}
        \end{subfigure} 

        \begin{subfigure}[b]{0.5\textwidth}
                \includegraphics[width=\textwidth]{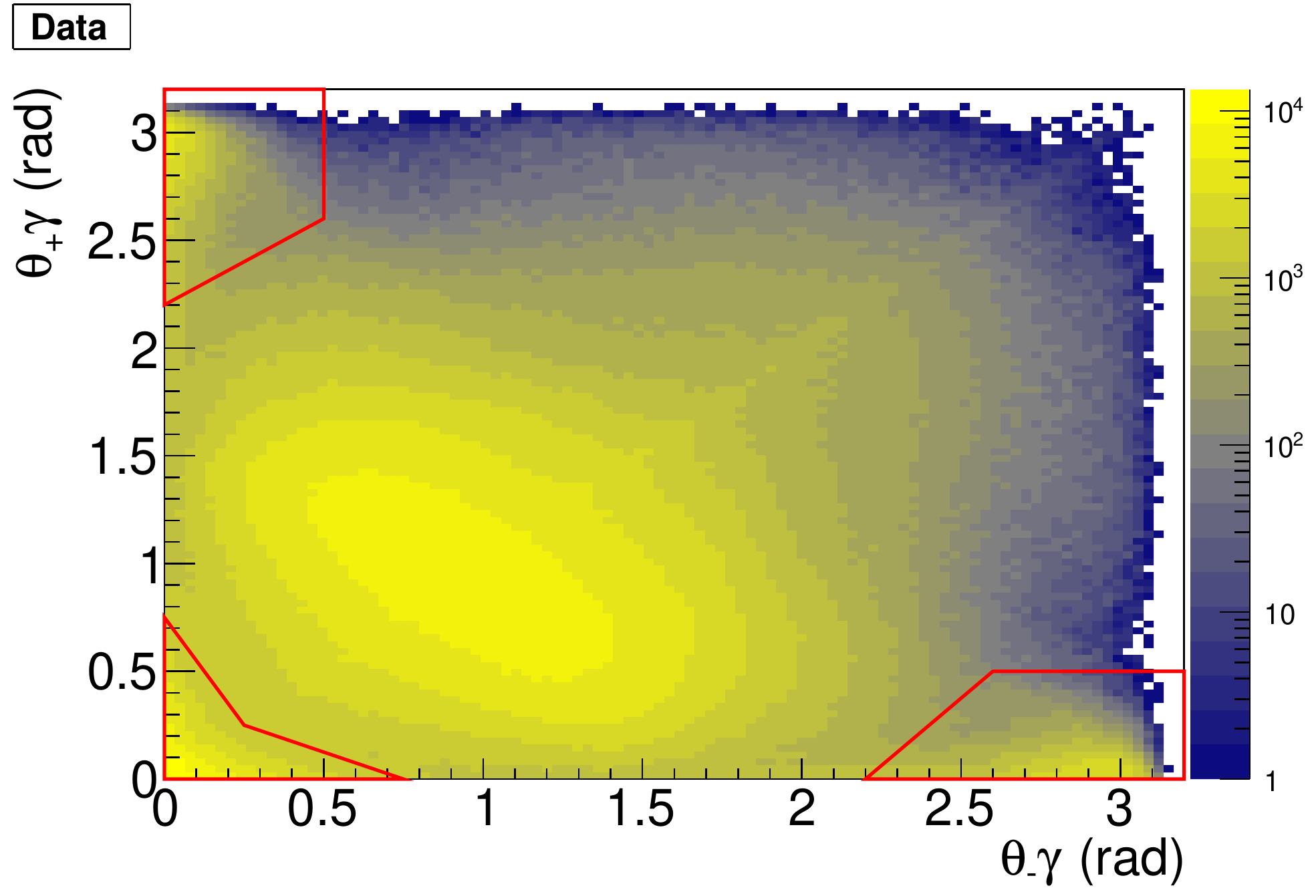}
        \end{subfigure}
\caption{(Color online) $\theta_{+\gamma}$ {\it vs} $\theta_{+\gamma}$
  angle plot. The three panels correspond to signal MC, Bhabha MC
  and the data.  The three regions in the corners with borders marked
  by red lines represent the Bhabha rejection cut applied in the
  analysis.\label{fig:graphcut}}
\end{figure*}

\begin{figure*}[htbp]
        \begin{subfigure}[b]{0.5\textwidth}
                \includegraphics[width=\textwidth]{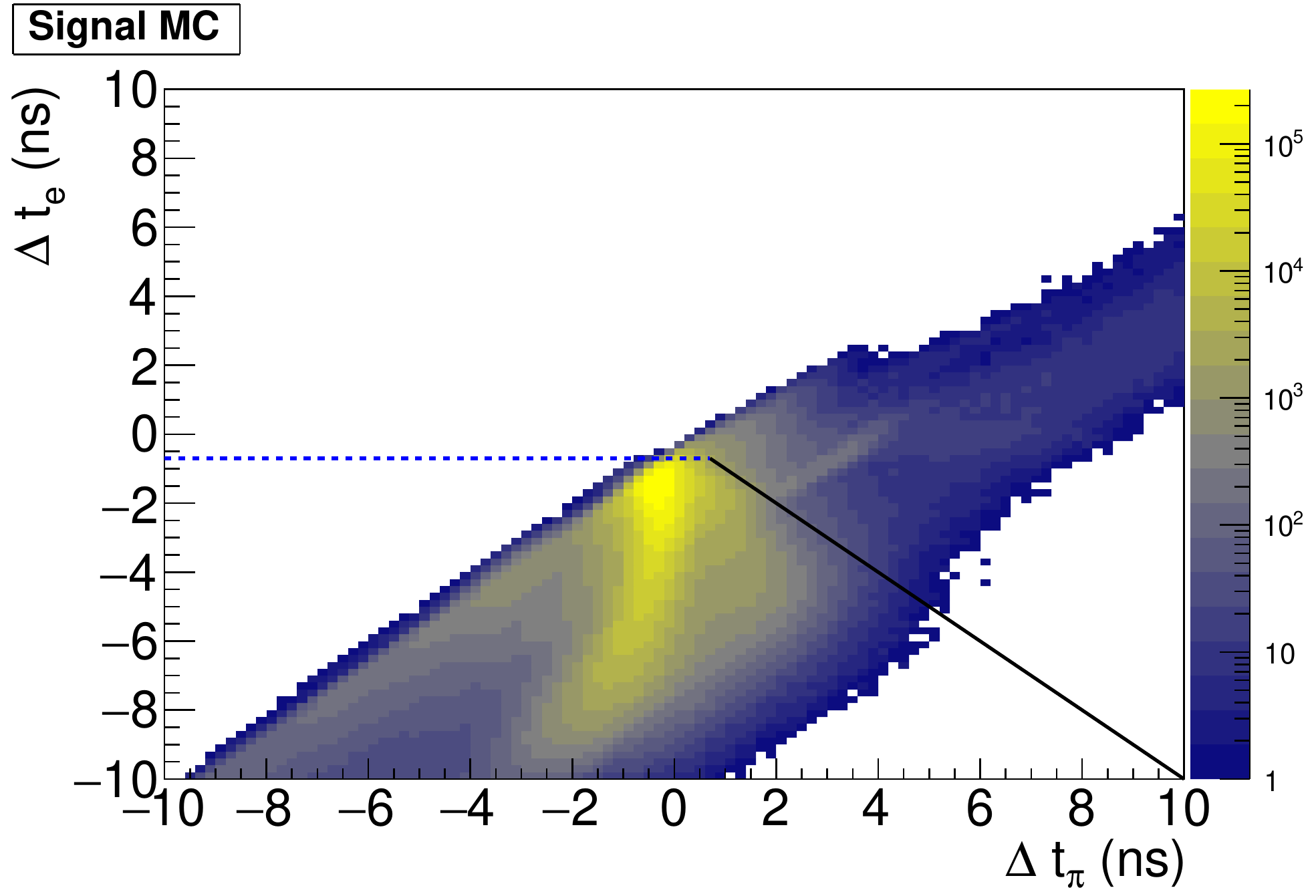}
        \end{subfigure}%
        ~ 
        \begin{subfigure}[b]{0.5\textwidth}
                \includegraphics[width=\textwidth]{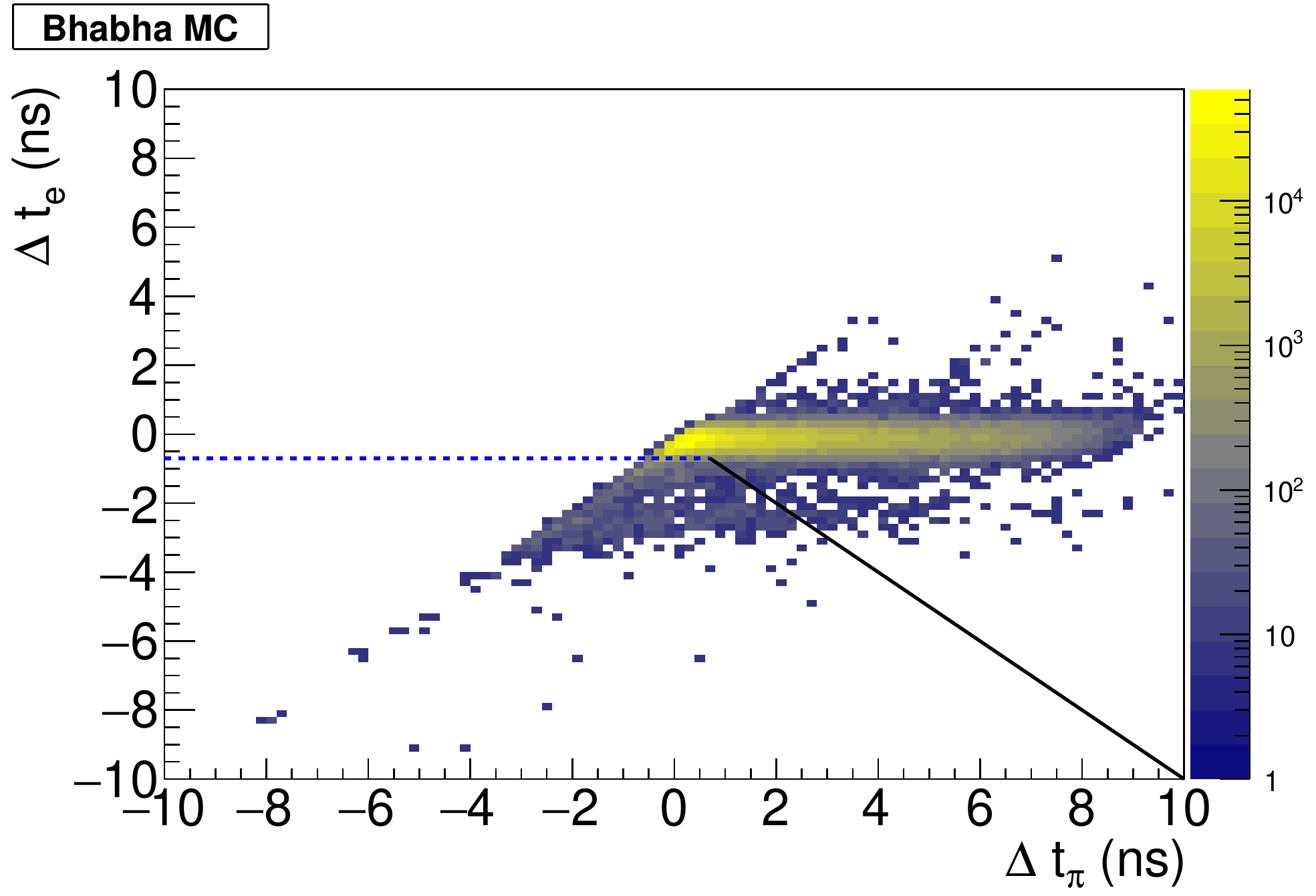}
        \end{subfigure} 

        \begin{subfigure}[b]{0.5\textwidth}
                \includegraphics[width=\textwidth]{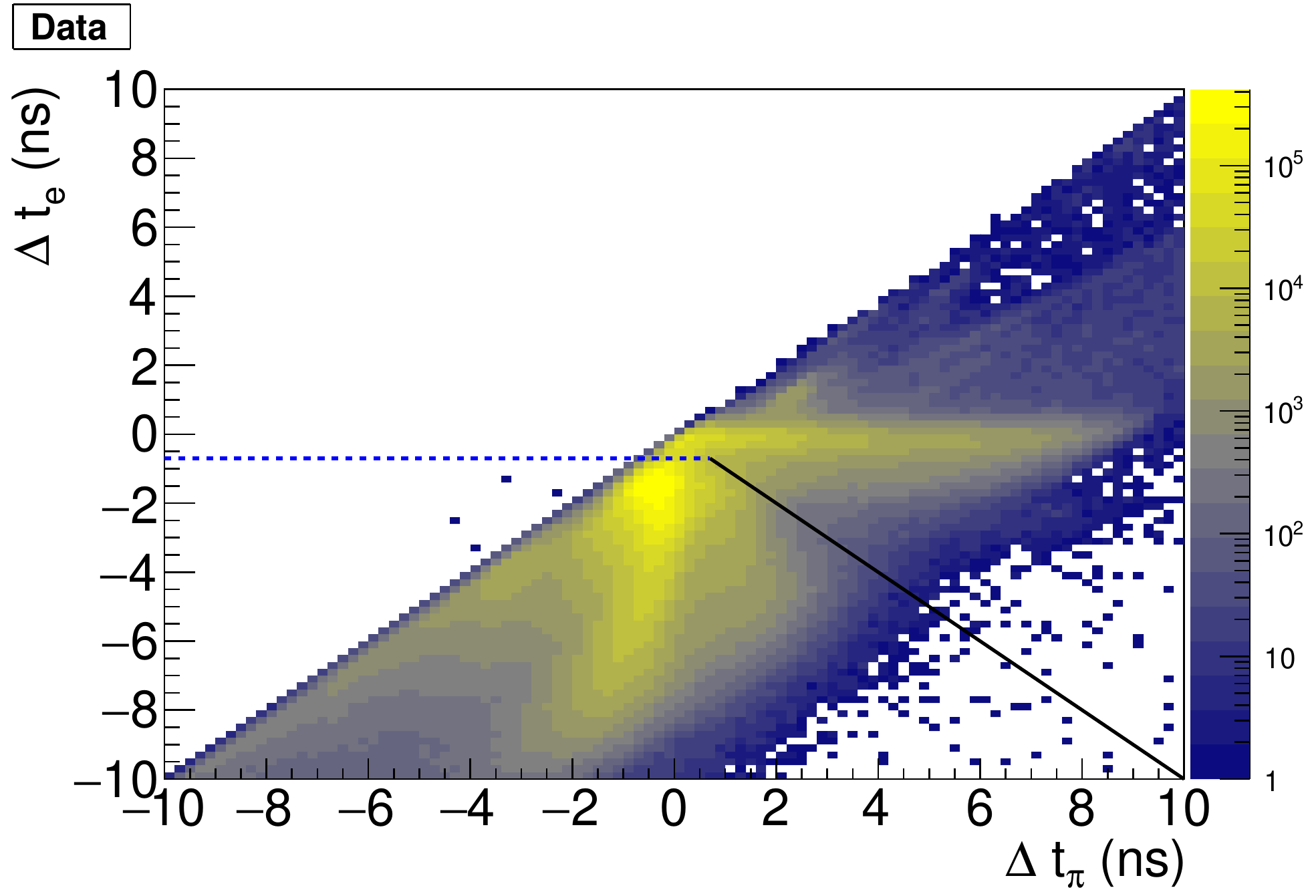}
        \end{subfigure}
\caption{(Color online) $\Delta t_e$ \emph{vs} $\Delta t_\pi$ plots for signal MC, Bhabha MC and the data.  Events above the blue (dotted) line or above the black (full) line are rejected. \label{fig:toffig}}
\end{figure*}

\begin{figure}[htbp]
\includegraphics[width=.5\textwidth]{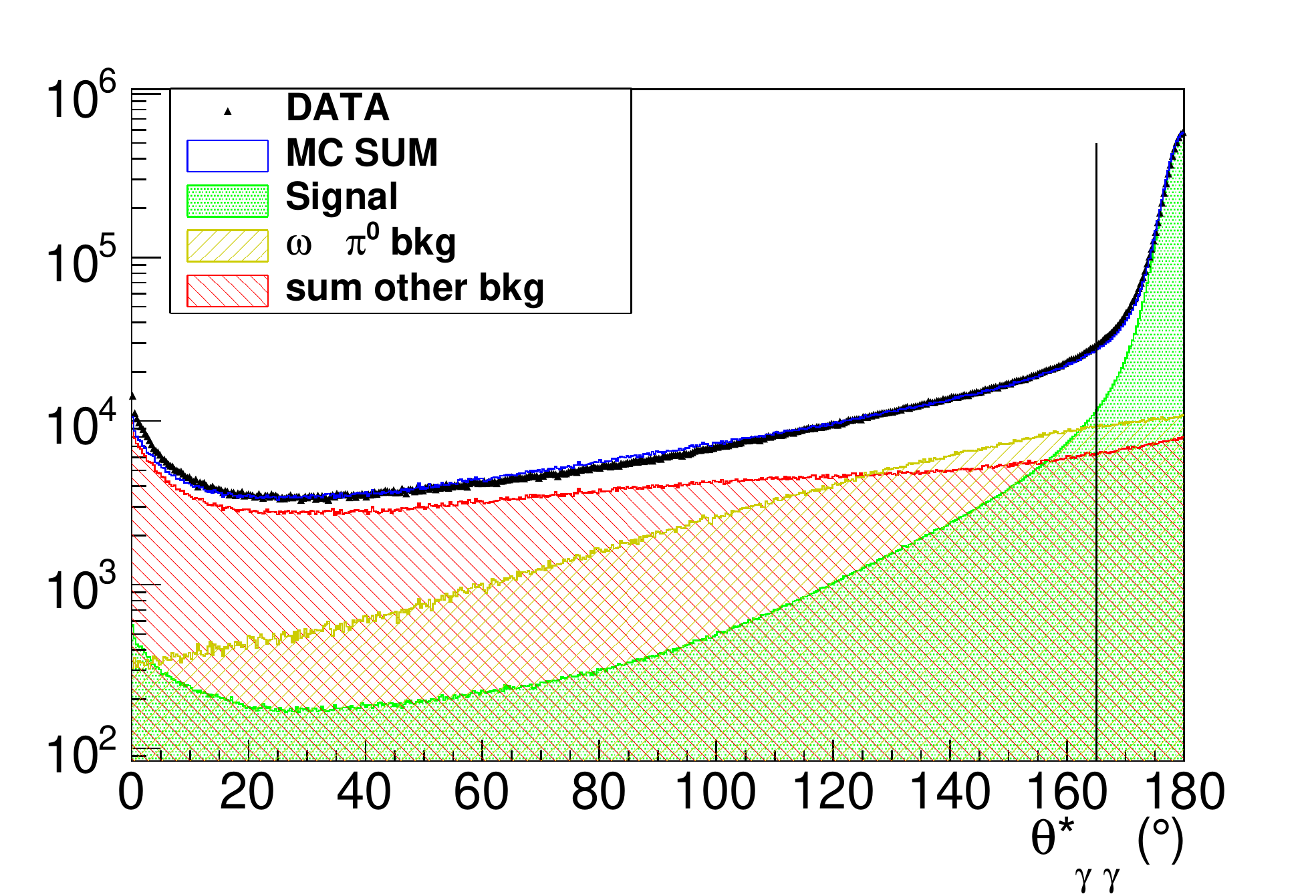}
\caption{(Color online) Azimuthal angle difference between the $\pi^0$ decay photons in
  the $\pi^0$ rest frame, $\theta^*_{\gamma\gamma}$, with the MC contributions scaled. The cut
   $\theta^*_{\gamma\gamma}>165^\circ$ is shown by the vertical
  line. \label{fig:opanglecut}}
\end{figure}

\begin{figure}[htbp]
\begin{center}
\includegraphics[width=.5\textwidth]{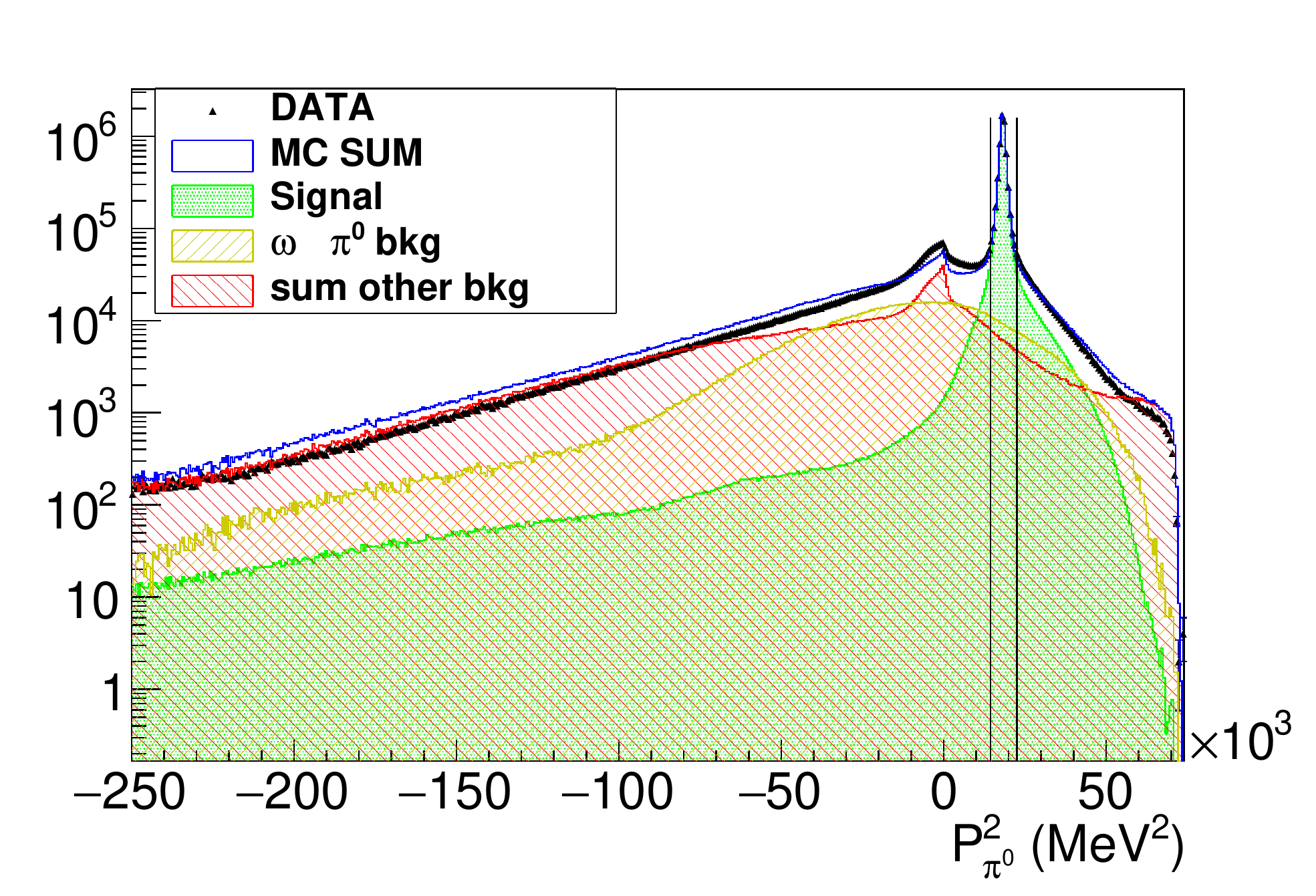}
\caption{(Color online) Missing mass squared, $P_{\pi^0}^2$, with the MC contributions scaled. The cut $\left||P_{\pi^0}|-m_{\pi^0}\right| < 15$ MeV is represented by 
   the two vertical
  lines. \label{fig:missmasssquared}}
\end{center}
\end{figure}

Two  tracks of
opposite curvature and three neutral clusters are expected in the final
state of the chain  $e^+ e^- \to
\phi \to \eta \gamma_\phi \to \pi^+ \pi^- \pi^0 \gamma_\phi \to \pi^+
\pi^- \gamma \gamma \gamma_\phi$.
Selection steps are listed below:
\begin{itemize}
\item  A candidate event has at least three prompt neutral clusters in the EMC. The clusters are required to have energy at least 10 MeV and polar angles $23^\circ < \theta< 157^\circ$, where $\theta$ is calculated from the distance of the cluster to the beam crossing point ($R_\text{cluster}$). The time of the prompt clusters should be within the time window for massless particles, $|t_\text{cluster} - R_\text{cluster}/c| < 5 \sigma(t)$, while neutral clusters do not have an associated track in the DC.
\item  At least one of the prompt neutral clusters has energy greater than 250 MeV. The highest energy cluster is assumed to originate from the $\gamma_\phi$ photon.
\item  The two tracks with smallest distance from the beam crossing, and with opposite curvature, are chosen. 
In the following these tracks are assumed to be due to charged pions. Discrimination against electron contamination from Bhabha scattering is achieved by means of Time Of Flight  as discussed in the following.
\item    $P_\phi$, the  four-momentum of the $\phi$ meson, is determined using the beam-beam energy
$\sqrt{s}$ and the $\phi$ transverse momentum measured in Bhabha scattering events for each run.
\item The  $\gamma_\phi$
direction is obtained from the position of the EMC cluster while its energy/momentum is calculated from the two body kinematics of the 
$\phi\to\eta\gamma_\phi$ decay: 
\[
E_{\gamma_\phi} = \frac{m_\phi^2 - m_\eta^2}{2\cdot\left( E_\phi - |{\bf p}_\phi| \cos \theta_{\phi,\gamma} \right)}
\] 
where $\theta_{\phi,\gamma}$ is the angle between the $\phi$ and the $\gamma_\phi$ momenta.
The four-momentum of the $\eta$ meson is then: $P_\eta = P_\phi - P_{\gamma_\phi}$.
\item The $\pi^0$ four-momentum is calculated from the missing four-momentum to $\eta$ and the charged pions: $P_{\pi^0} = P_\eta - P_{\pi^+}- P_{\pi^-} $.
\item To reduce the Bhabha scattering background, the following two cuts are applied:
\begin{itemize}
\item 
a cut in the ($\theta_{+\gamma}$,$\theta_{-\gamma}$) plane as shown in  Fig.~\ref{fig:graphcut}, where $\theta_{+\gamma}(\theta_{-\gamma})$ is the angle between the $\pi^+(\pi^-)$ and the closest photon from $\pi^0$ decay.
\item 
a cut in the $(\Delta t_e,\Delta t_\pi)$ plane as shown in
Fig.~\ref{fig:toffig}, to discriminate electrons from pions, where
$\Delta t_e$, $\Delta t_\pi$ are calculated for tracks which have an
associated cluster, $\Delta t_{e / \pi}\equiv t_{\text{track}_{e /
    \pi}} - t_\text{cluster}$, where $ t_{\text{track}_{e / \pi}}$, is
the expected arrival time to EMC for $e/\pi$ with the measured
momentum, and $ t_\text{cluster}$ the measured time of the EMC
cluster.
\end{itemize}
\item To improve the agreement between simulation and data, a correction for the relative yields of: (i) $e^+e^-\to \omega \pi^0$, and (ii) sum of all other backgrounds, with respect to the signal is applied. The correction factors are obtained from a fit to the distribution of the azimuthal angle between the $\pi^0$ decay photons, in the $\pi^0$ rest frame, $\theta_{\gamma\gamma}^*$ (Fig.~\ref{fig:opanglecut}).  The uncertainties of the correction factors are evaluated by comparing the corresponding fit to the distribution of the missing mass squared, $P_{\pi^0}^2$ (Fig.~\ref{fig:missmasssquared}). 
\item To further reduce the background contamination, two more cuts are applied:
\begin{itemize}
\item $\theta_{\gamma\gamma}^*>165^\circ$, see Fig.~\ref{fig:opanglecut};
\item $\left||P_{\pi^0}|-m_{\pi^0}\right|<15$ MeV, see Fig.~\ref{fig:missmasssquared};
\end{itemize}
\end{itemize}
The overall signal efficiency is 37.6\% at the end of the analysis
chain and the signal to background ratio is 133.

\begin{figure*}[htbp]
\centering
        \begin{subfigure}[b]{0.45\textwidth}
                \includegraphics[width=\textwidth]{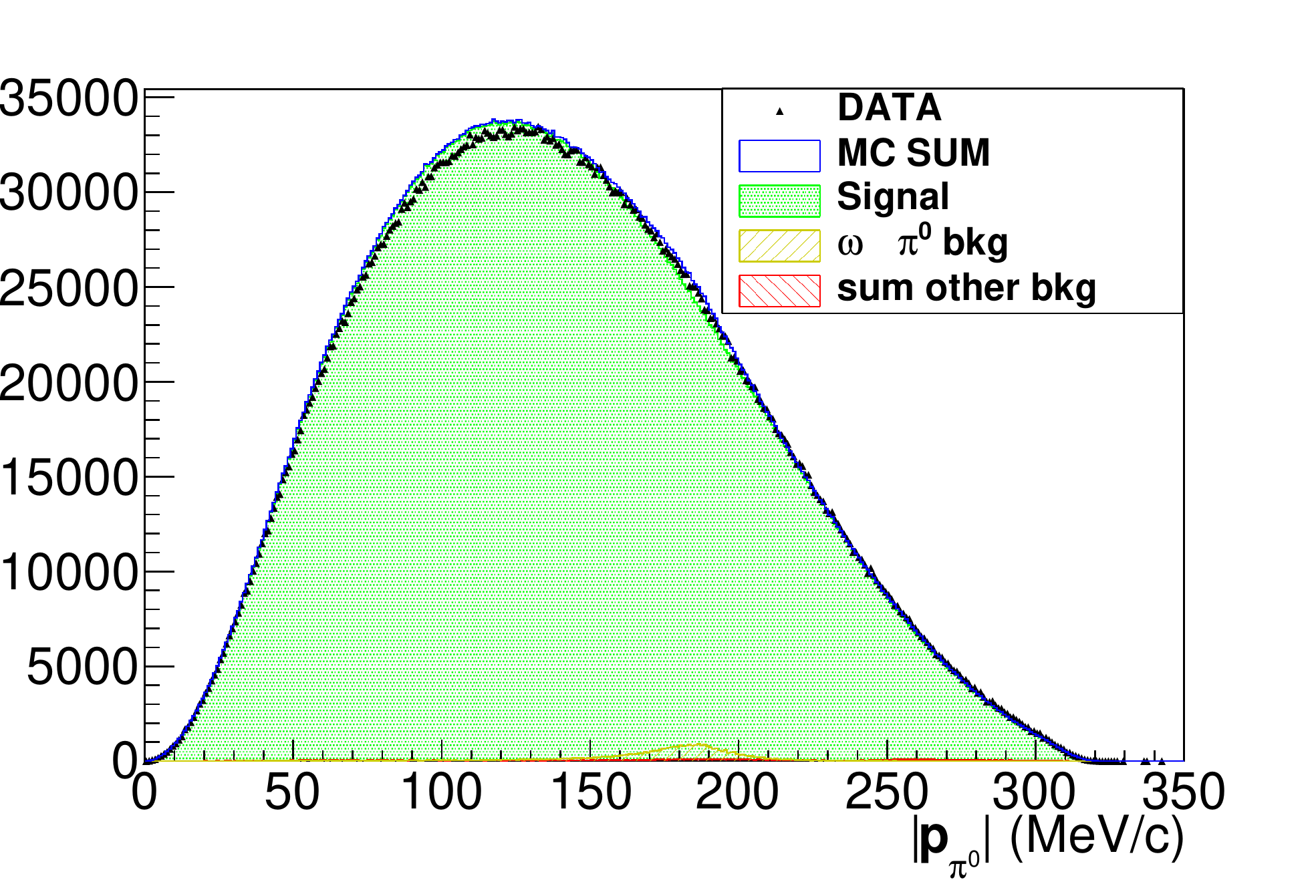}
        \end{subfigure}%
        ~ 
        \begin{subfigure}[b]{0.45\textwidth}
                \includegraphics[width=\textwidth]{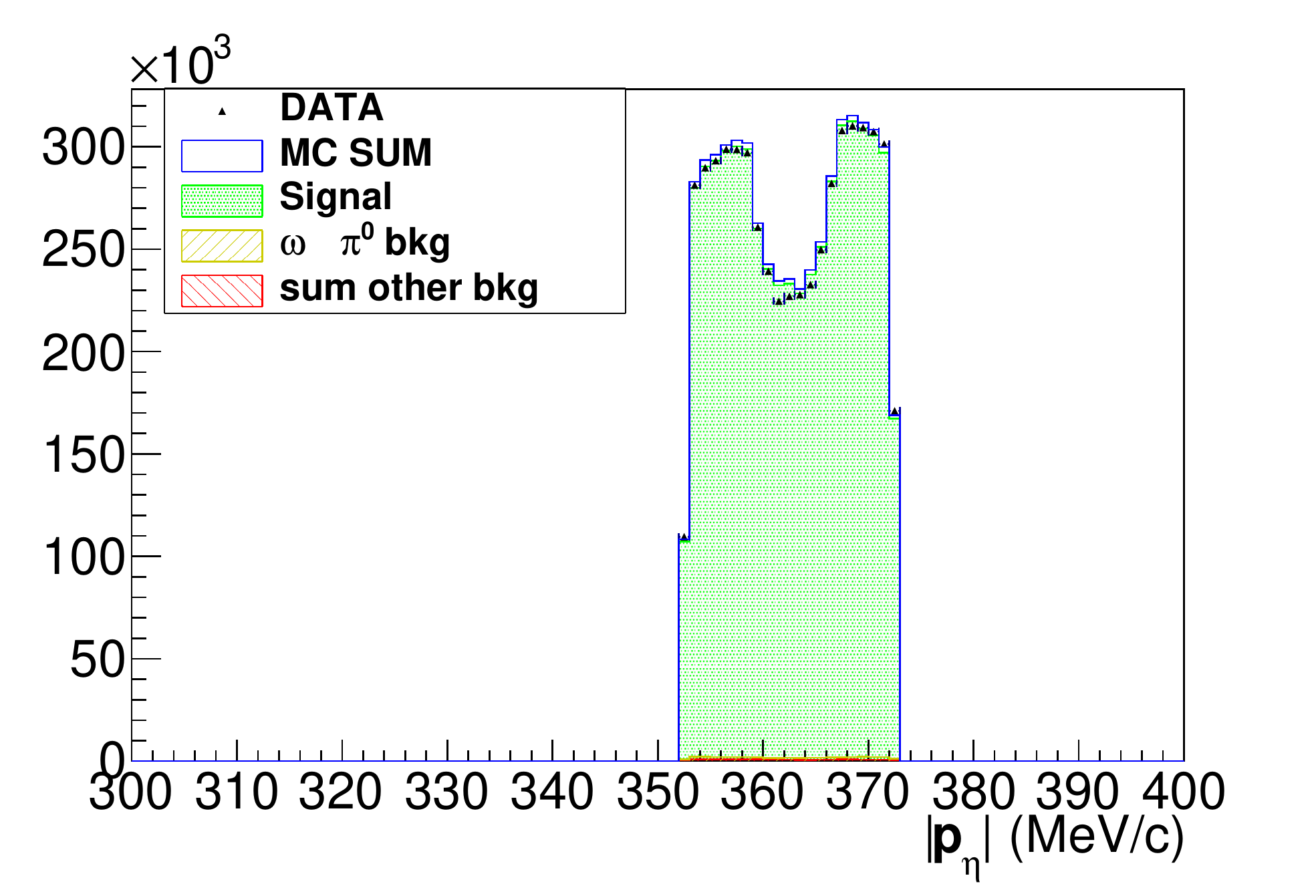}
        \end{subfigure}
\caption{Distribution of the reconstructed momentum  of $\pi^0$ (left) and  $\eta$ (right) for  the data and
 MC. \label{fig:neutralsmom}}
\end{figure*}

As can  be seen in Figs.~\ref{fig:opanglecut}, \ref{fig:missmasssquared} and \ref{fig:neutralsmom} the 
agreement of simulation with
the experimental data is good.

\begin{figure*}[htbp]
\centering
        \begin{subfigure}[b]{0.45\textwidth}
                \includegraphics[width=\textwidth]{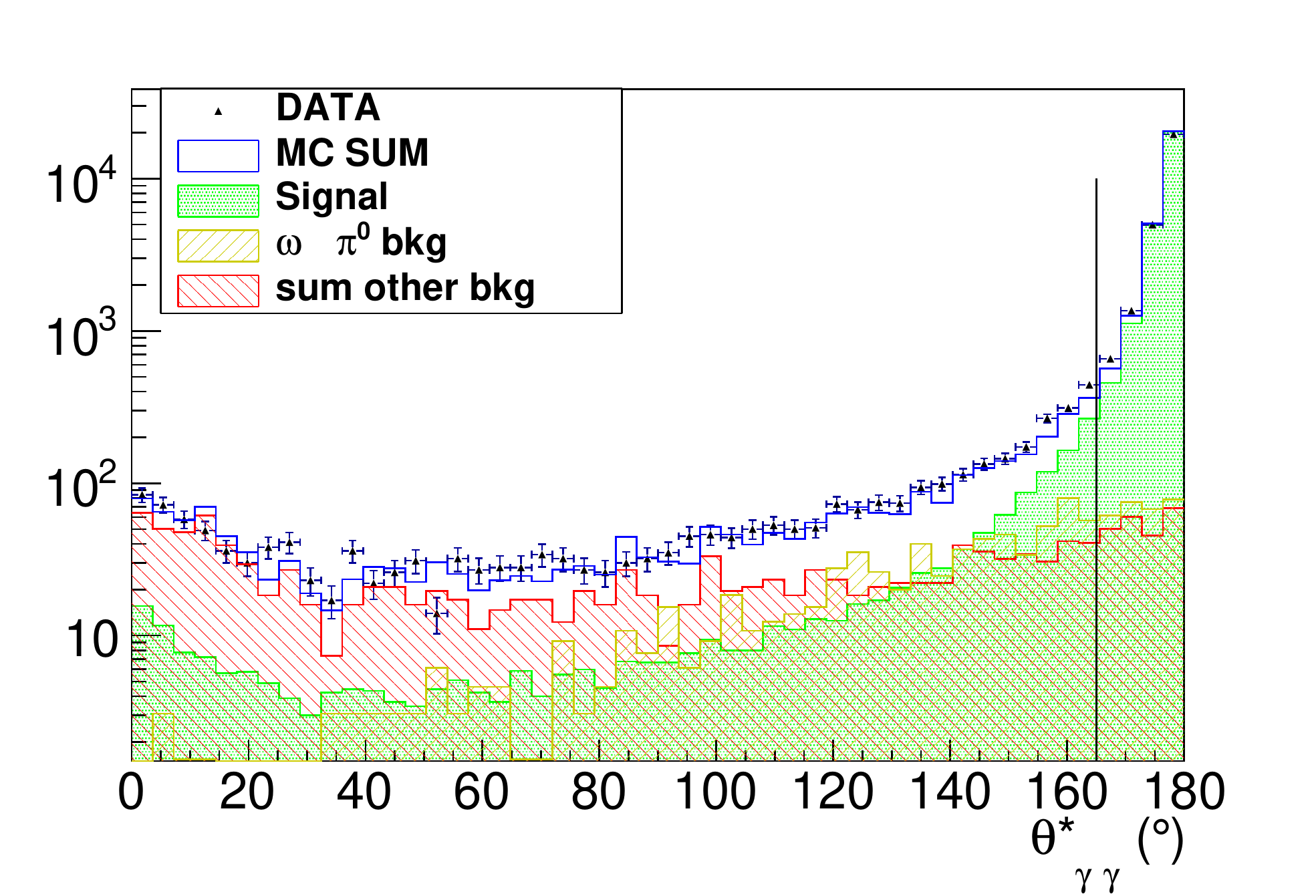}
        \end{subfigure}%
        ~ 
        \begin{subfigure}[b]{0.45\textwidth}
                \includegraphics[width=\textwidth]{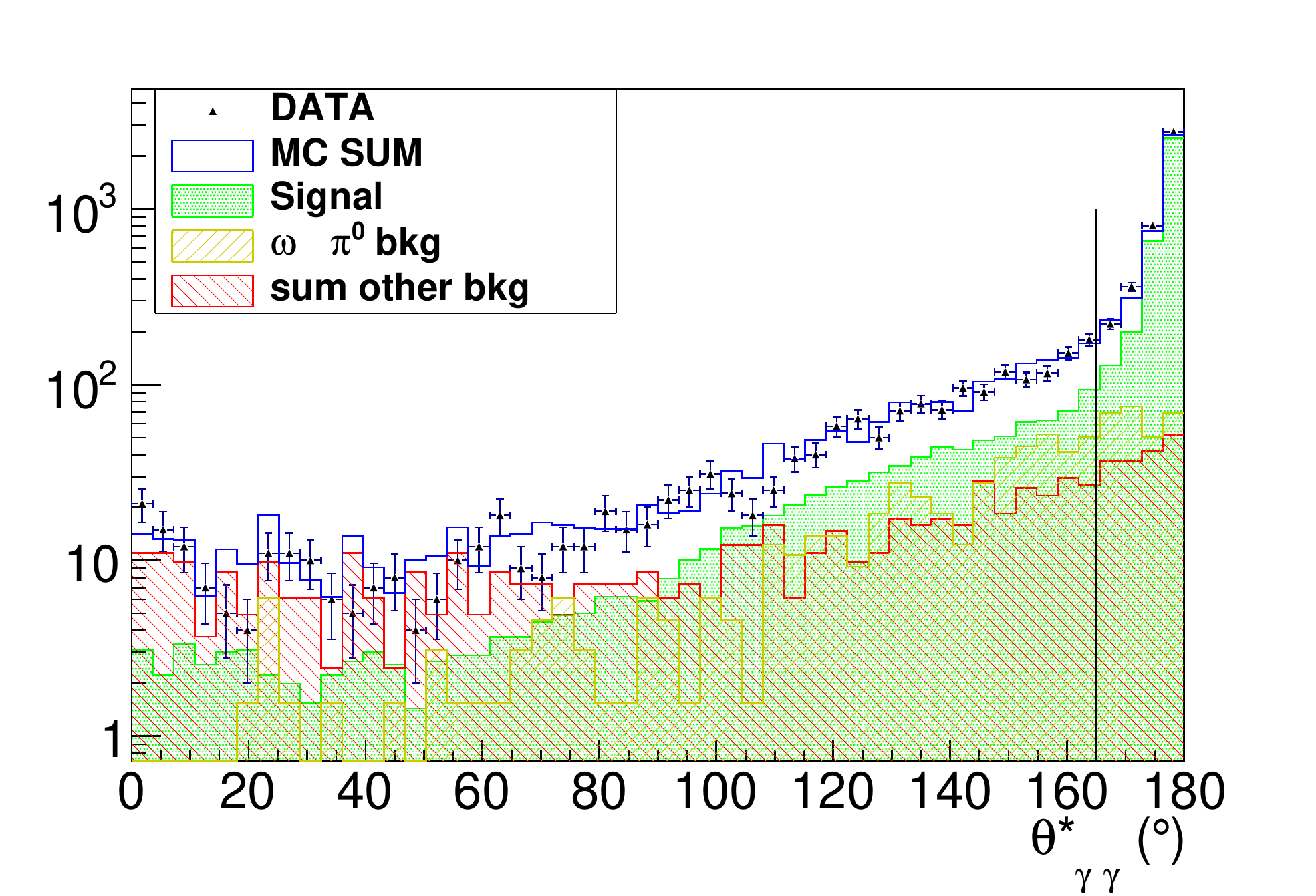}
        \end{subfigure}
        
        \begin{subfigure}[b]{0.45\textwidth}
                \includegraphics[width=\textwidth]{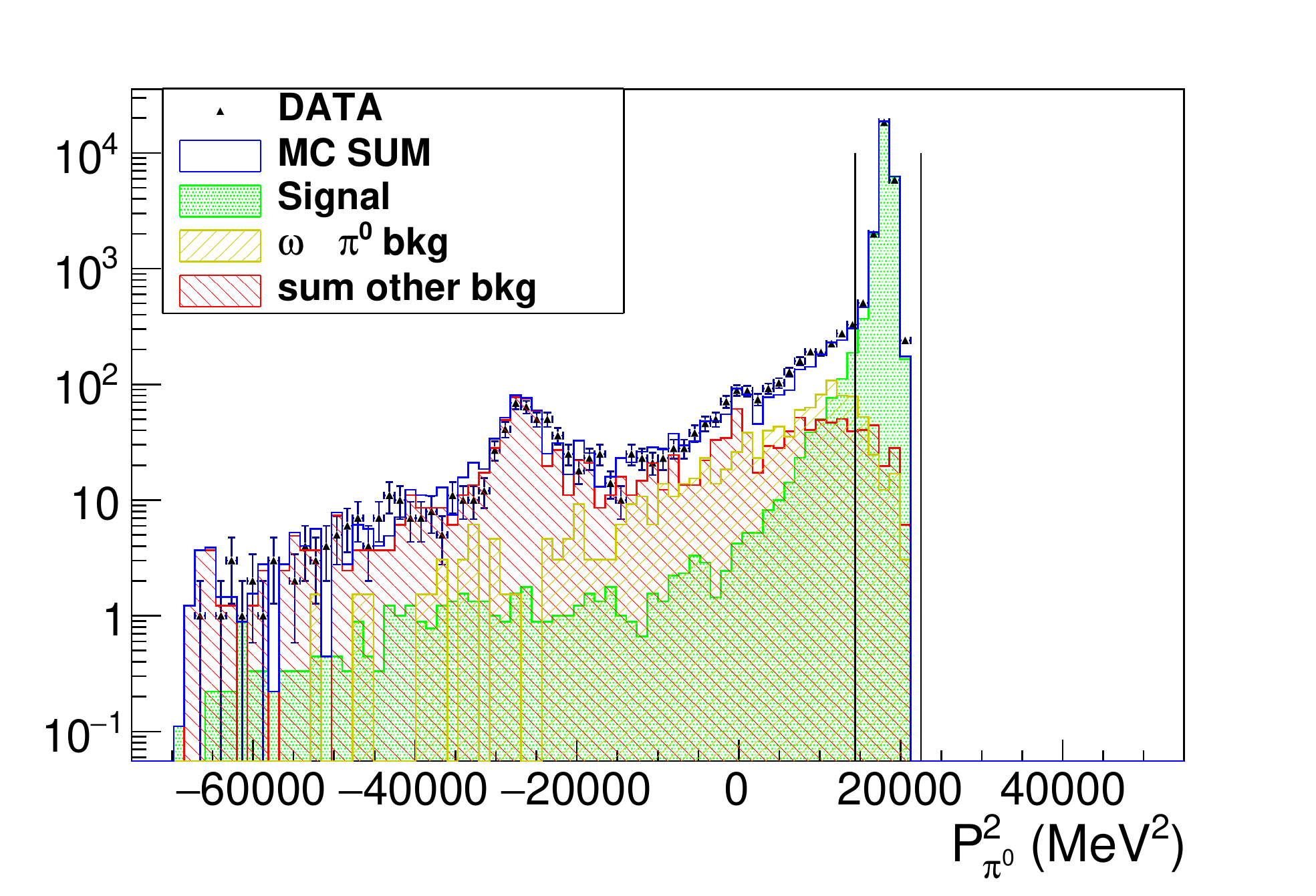}
        \end{subfigure} 
        ~ 
        \begin{subfigure}[b]{0.45\textwidth}
                \includegraphics[width=\textwidth]{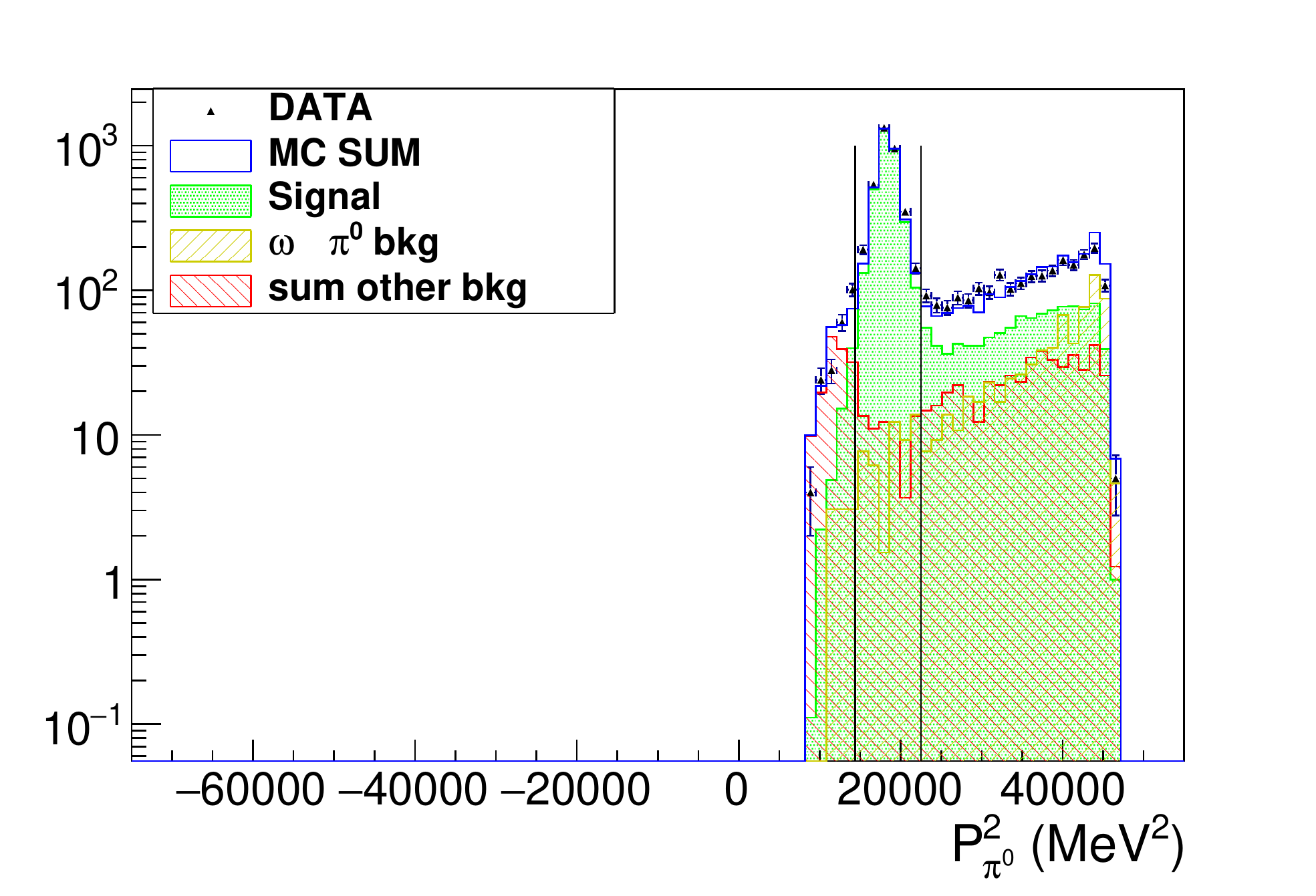}
        \end{subfigure}
\caption{(Color online) Top:
  $\theta_{\gamma\gamma}^*$ angle distribution  with the MC contributions scaled; the
  selected region is at the right of the vertical line.  Bottom:
  missing mass squared,$P_{\pi^0}^2$, with the MC contributions
  scaled. The selected region is between the vertical
  lines. Left/right: bin of the Dalitz plot with the largest/smallest
  number of entries, corresponding to $(X,Y)=(0.000,-0.850)$ and
  $(X,Y)=(-0.065,0.750)$, respectively.
 \label{fig:opang_mm_binbybin}}
\end{figure*}

\section{Dalitz Plot}
For the Dalitz plot, a two dimensional histogram representation is
used.  The bin width is determined both by the resolution in the $X$
and $Y$ variables and the number of events in each bin, which should
be large enough to justify $\chi^2$ fitting.  The resolution of the
$X$ and $Y$ variables is evaluated with MC signal simulation. The
distribution of the difference between the true and reconstructed
values is fit with a double Gaussian. The standard deviations of the
narrower Gaussians are $\delta_X = 0.021$ and $\delta_Y = 0.032$.  The
range $(-1,1)$ for the $X$ and $Y$ variables was divided into 31
and 20 bins, respectively.  Therefore the bin widths correspond to
approximately three standard deviations. The minimum bin content is $
3.3 \cdot 10^3 $ events. Fig.~\ref{fig:opang_mm_binbybin} shows the
distributions of the $\theta_{\gamma\gamma}^*$ and the $P_{\pi^0}^2$
variables for two bins in the Dalitz plot, one with the largest
content and one with the smallest. As can be seen, the signal and the
background are well reproduced by the simulation.


\begin{figure}[htbp]
\centering
\includegraphics[width=0.5\textwidth]{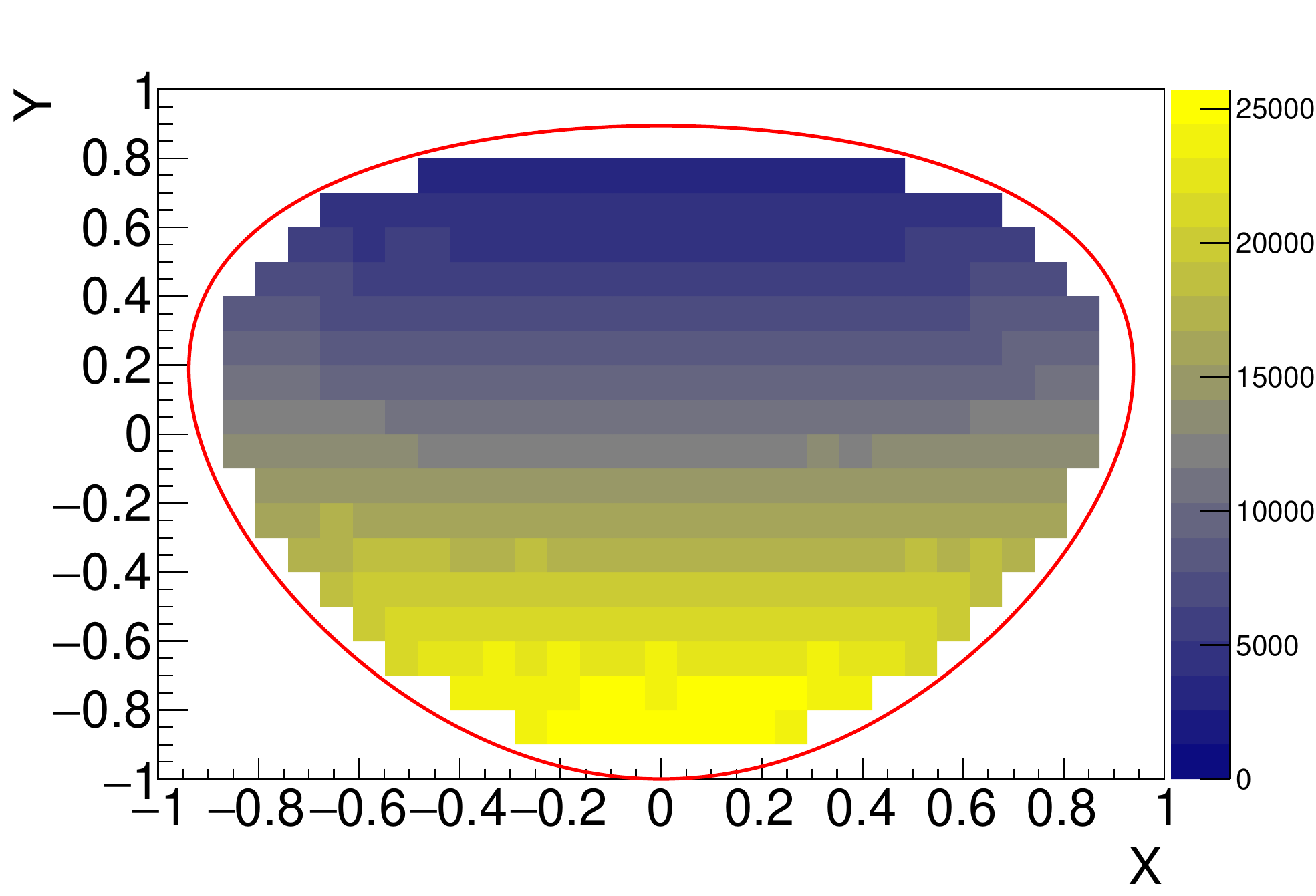}
\caption{(Color online) The experimental background subtracted Dalitz
  plot distribution represented by the two dimensional histogram with
  371 bins. Only bins used for the Dalitz parameter fits are
  shown. The physical border is indicated by the red
  line. \label{fig:dpborder}}
\end{figure}

Fig.~\ref{fig:dpborder} shows the experimental Dalitz plot
distribution after background subtraction, which is fit to the
amplitude expansion from Eq.~(\ref{eq:DPamplitude}) to extract the
Dalitz plot parameters. Only $n=371$ bins which are fully inside the kinematic
boundaries are used and there are $\sim 4.7 \cdot 10^6$
entries in the background subtracted Dalitz plot.

The fit is performed by minimizing the $\chi^2$ like function
\begin{equation}
\chi^2 = \sum_{i=1}^{n} \left( \frac{N_i - \sum_{j=1}^{n_T}  S_{ij} N_{T,j}}{\sigma_i}    \right)^2 \label{eq:chi2}
\end{equation}
where:
\begin{itemize}
\item $N_{T,j} = \int |A(X,Y)|^2 dPh(X,Y)_j$, with $|A(X,Y)|^2$ given by Eq.~(\ref{eq:DPamplitude}). The integral is over $X$ and $Y$ in the allowed phase space for bin $j$. The sum over $j$ bins includes all Dalitz plot bins at least partly inside the physical border, $n_T$. 
\item $ N_i = N_{data,i} - \beta_{1} B_{i1}- \beta_{2} B_{i2}$\label{bkgsub} is the background subtracted content of Dalitz plot bin $i$, where $\beta_{1,2}$ are the scaling factors, $B_{i1}$ is the $\omega \pi^0$ background in the bin $i$ and $B_{i2}$ is the same for the remaining background.
\item  $S_{ij}$ is the acceptance and smearing matrix from bin $j$ to bin $i$ in the Dalitz plot. It is determined from signal MC by  
$S_{ij} = {N_{rec,i;gen,j}}/{N_{gen,j}}$, 
where $N_{rec,i;gen,j}$ denotes the number of events reconstructed in bin $i$ which were generated in bin $j$ and $N_{gen,j}$ denotes the total number of events generated in bin $j$. 
\item $\sigma_i^2= \sigma_{N_i}^2 + \sigma_{S_{ij}}^2 $ is the error in bin $i$, with  
$\sigma_{S_{ij}}^2 = \sum_{j=1}^{n_T} N_{T,j}^2 \cdot {S_{ij} \cdot (1-S_{ij})}/{N_{gen,j}}$. 
\end{itemize}

The input-output test of the fit procedure was performed using signal
MC generated with the same statistics as the experimental data. The
extracted values for the parameters were within one standard deviation
with respect to the input.


\begin{table*}[htbp]
\caption[]{Results for the Dalitz plot parameter fits. Fit $\#4$
  includes the $g$ parameter, while fit $\#3$, with $g = 0$, can be
  directly compared to previous 
  results. The fits $\#5$ and $\#6$ use the acceptance corrected data
  (see Appendix~\ref{sec:APPA}).\label{tab:dpfit_results}}
\begin{tabular}{lcccccccc}
\hline\hline
Fit/set\# & $a$  & $b\cdot 10$ &  $d\cdot 10^{2}$ &  $f\cdot 10$  & $g\cdot 10^{2}$& $c,e,h,l$ & $\chi^2/$dof & Prob\\\hline
(1) & $-1.095\pm0.003$ &$ 1.454\pm0.030$ & $8.11\pm0.32$ & $1.41\pm0.07$  & $-4.4\pm0.9$ & free&$354/361$ & 0.60 \\ 
(2) & $-1.104\pm0.002$ &$ 1.533\pm0.028$ & $6.75\pm0.27$ & $0$ & $0$ &$0$ &$1007/367$ & $0$  \\
(3) & $-1.104\pm0.003$ &$ 1.420\pm0.029$ & $7.26\pm0.27$ & $1.54\pm0.06$  & $0$ & $0$&$385/366$ & 0.24 \\ 
(4) & $-1.095\pm0.003$ &$ 1.454\pm0.030$ & $8.11\pm0.33$ & $1.41\pm0.07$  &$-4.4\pm0.9$ &$0$ &$360/365$   &0.56    \\
\hline
(5)& $-1.092\pm0.003$ &$ 1.45\pm 0.03$ & $8.1\pm0.3$ & $1.37\pm0.06$ & $-4.4\pm0.9$&$0$ &$369/365$& 0.43  \\
(6)& $-1.101\pm0.003$ &$1.41\pm 0.03 $ & $7.2\pm0.3$ &$1.50\pm0.06$  & $0$ &$0$ &$397/366$& 0.13 \\
\hline\hline
\end{tabular}
\end{table*}
The fit has been performed using different choices of the free
parameters in Eq.~(\ref{eq:DPamplitude}), with the normalization $N$
and the parameters $a$, $b$ and $d$ always let free.  The main fit
results are summarized in Tab.~\ref{tab:dpfit_results}. The first row
(set \#1) includes all parameters of the cubic expansion,
Eq.~(\ref{eq:DPamplitude}). The fit values of the charge conjugation
violating parameters $c,e,h$ and $l$ are consistent with zero ($c
=(4.3\pm3.4)\cdot 10^{-3} $, $e=(2.5\pm3.2)\cdot 10^{-3}$,
$h=(1.1\pm0.9)\cdot 10^{-2}$, $l=(1.1\pm6.5)\cdot 10^{-3} $) and are
omitted from the table.  Therefore in all remaining fits the charge
conjugation violating parameters $c,e,h$ and $l$ are set to zero.  The
result \#2 with $f=g=0$ demonstrates that it is not possible to
describe the distribution with only quadratic terms.  The fits
including in addition the $f$ parameter improves the accuracy
largely providing a good description. A complementary
test with $f$ fixed to zero and free $g$ parameter leads to much worse
$\chi^2$ value. This observation is consistent with the recent
experimental results that used the same free parameters, in particular
the KLOE(08) analysis. In the set \#4 which includes both $f$ and $g$
parameters in the fit, the $g$ parameter differs from zero at the
$4.9\sigma$ level and the fit probability is as good as for the 
fit \#1.  For completeness in the further discussions we
include also set \#3 with $g=0$, since it enables direct comparison to
the previous experiments (KLOE(08), WASA(14) and BESIII(15)). The
correlation matrices for fits \#3 and \#4 are:
\begin{center}
\begin{tabular}{ c| l l l }
 &  $b$ &$d$&$f$\\
\hline
$a$  &$-0.269$& $-0.365$& $-0.832$\\
$b$  &  & $+0.333$& $-0.139$\\
$d$  & &   &$+0.089$\\
\end{tabular}
\vskip.5cm
\begin{tabular}{ c |l l l l }
 & \multicolumn{1}{c}{$b$} &\multicolumn{1}{c}{$d$}&\multicolumn{1}{c}{$f$}&\multicolumn{1}{c}{$g$}\\   
\hline
$a$ &  $-0.120$&  $+0.044$& $-0.859$& $-0.534$\\
$b$ &   & $+0.389$& $-0.201$& $-0.225$\\
$d$ &  &  &$-0.160$& $-0.557$\\
$f$ &  & &  & $+0.408$.\\
\end{tabular}

\end{center}

\begin{figure}[p]
\includegraphics[width=0.5\textwidth]{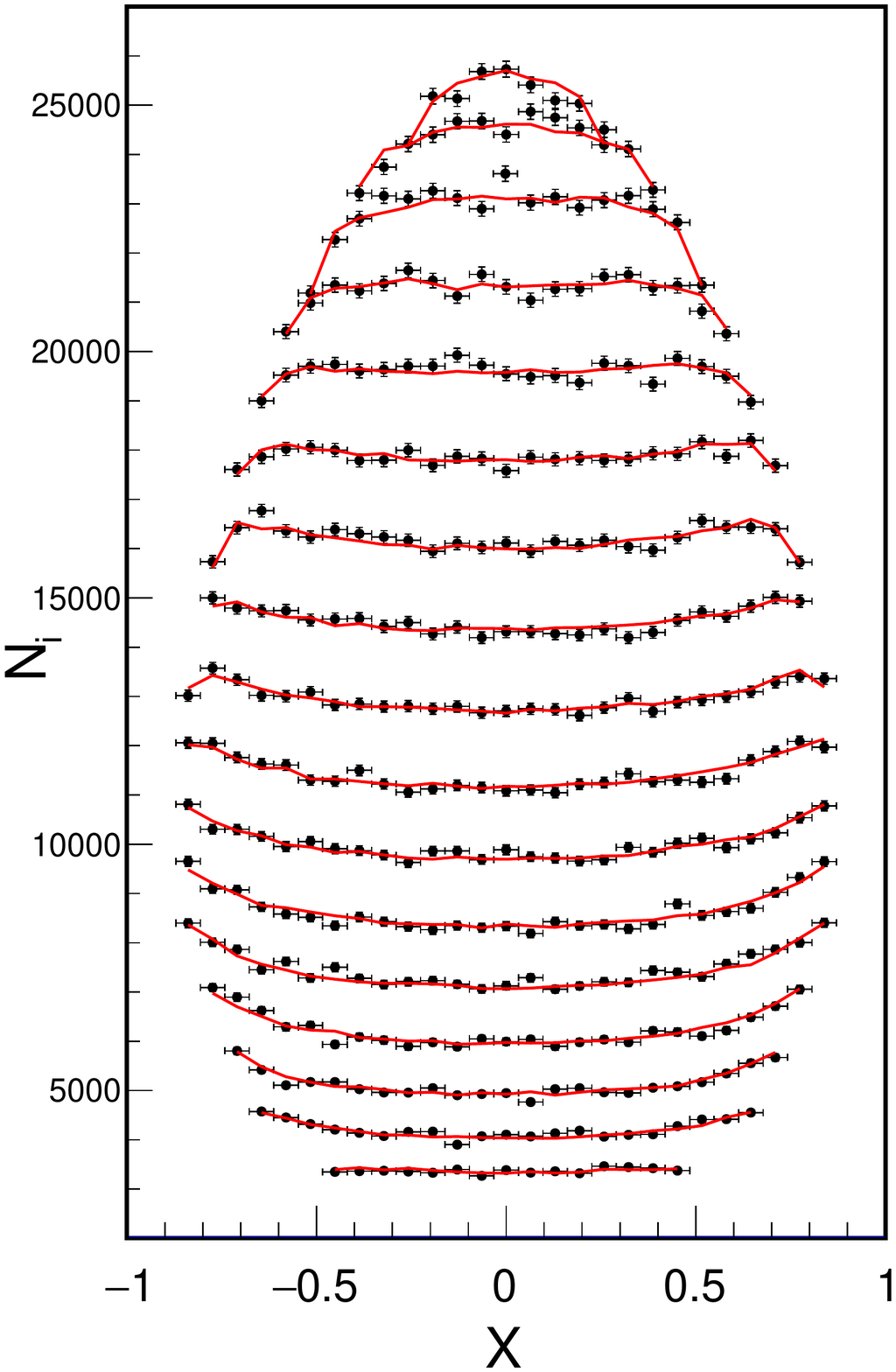}
\caption{(Color online) The experimental background subtracted Dalitz plot data, $N_i$, (points with errors), compared to set \#4 fit results (red lines connecting bins with the same $Y$ value). The row with lowest $N_i$ 
values corresponds to the highest $Y$ value ($Y=+0.75$). \label{fig:slicesinY}}
\end{figure}
\begin{figure}[htbp]
\centering
                \includegraphics[width=0.5\textwidth]{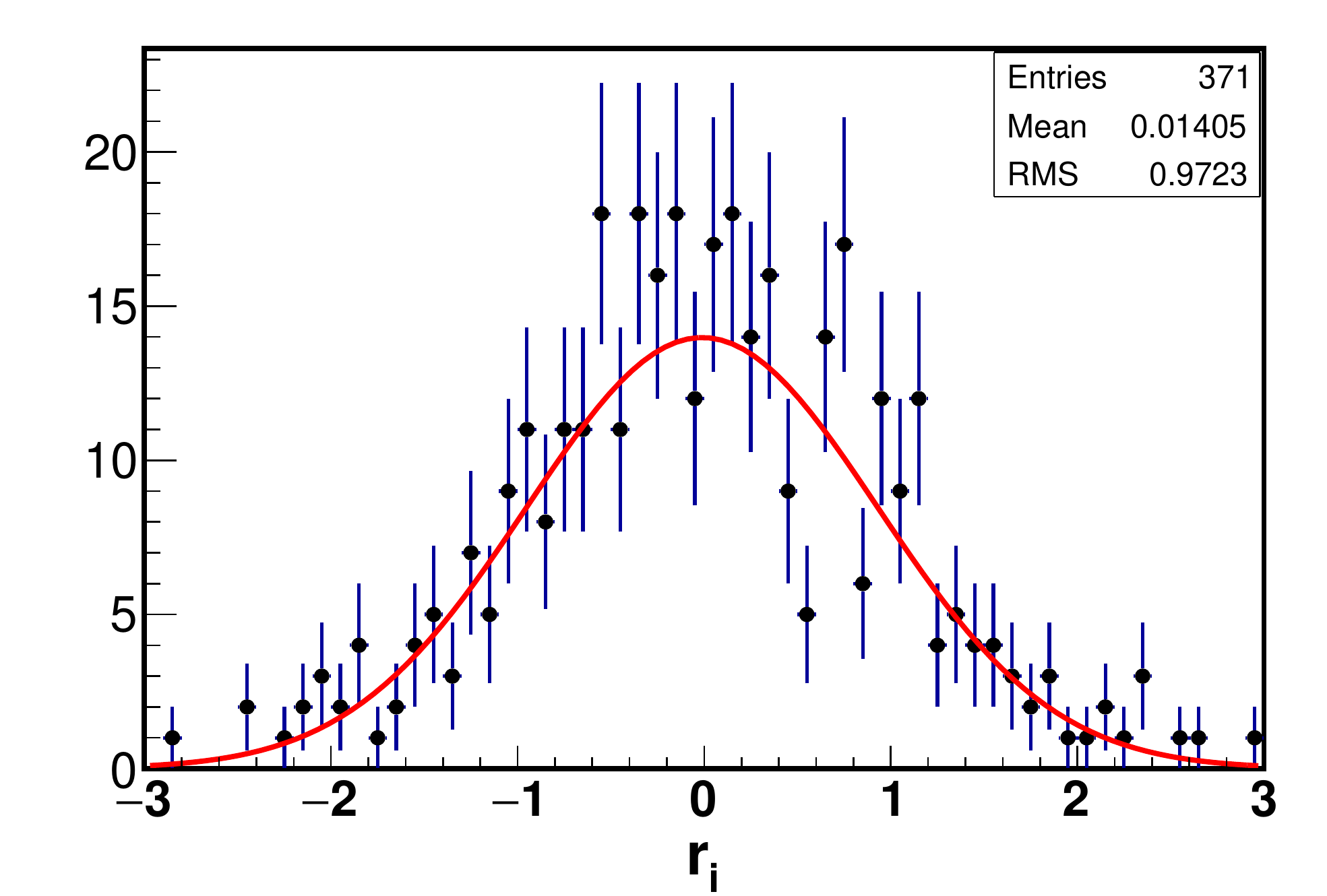}
\caption{(Color online) Distribution of the normalized residuals,
  $r_i$, for  fit \#4. \label{fig:normresiduals}}
\end{figure}

 The fit \#4 is compared to the background subtracted Dalitz plot
 data, $N_i$, in Fig.~\ref{fig:slicesinY}. The red lines represent the
 fit result and correspond to separate slices in the $Y$ variable.
 Fig.~\ref{fig:normresiduals} shows the distribution of the normalized
 residuals for the fit \#4: $r_i=(N_i - \sum_{j=1}^{n} S_{ij}
 N_{T,j})/\sigma_i$. The location of the residuals $r_i > 1$ and
 $r_i < -1$ on the Dalitz plot is uniform.    The fits $\#5$ and $\#6$ use
 the acceptance corrected data (see Appendix~\ref{sec:APPA}).

\section{Asymmetries}
While the extracted Dalitz plot parameters are consistent with charge
conjugation symmetry, the unbinned integrated charge asymmetries
provide a more sensitive test. The left-right ($A_{LR}$), quadrant
($A_Q$) and sextant ($A_S$) asymmetries are defined in
Ref.~\cite{layter72_asym}. The same background subtraction is applied
as for the Dalitz plot parameter analysis. For each region in the
Dalitz plot used in the calculation of the asymmetries, the acceptance
is calculated from the signal MC as the ratio between the number of
the reconstructed and the generated events.  The yields are then
corrected for the corresponding efficiency. The procedure was tested
using signal MC generated with the same statistics as the experimental
data.  

\section{Systematic Checks}
To quantify and account for systematic effects in the results, several
checks have been made.  
\begin{itemize}
\item Minimum photon energy cut (EGmin) is changed from 10 MeV to 20
  MeV (for comparison
the EMC energy resolution varies from  60\% to 40\% for this 
energy range). The systematic error is taken as half of the difference.
\item Background subtraction (BkgSub) is checked by determining the
  background scaling factors for each bin (or region for the asymmetries) of the Dalitz plot
  separately.  With the same method as for the whole data sample, using the
  $\theta_{\gamma\gamma}^*$ and $P_{\pi^0}^2$ distributions, 
  background scaling factors are determined for each bin (or region). The
  systematic error is taken as half the difference with the standard
  result.
\item Choice of binning (BIN) is tested by varying number of bins of
  the Dalitz plot. For $X$ and $Y$ simultaneously, the bin width is
  varied from $\sim 2 \delta_{X,Y}$ to $\sim 5 \delta_{X,Y}$, in total
  10 configurations. The systematic uncertainty is given by the
  standard deviation of the results. 
\item $\theta_{+\gamma},\theta_{-\gamma}$ cut: the areas of the three
  zones shown in Fig.~\ref{fig:graphcut} were simultaneously varied by
  $\pm 10 \%$.
\item $\Delta t_e,\Delta t_\pi$ cut: the offsets of the horizontal and
  diagonal lines shown in Fig.~\ref{fig:toffig} were varied by $\pm
  0.22$ ns and $\pm0.21$ ns, respectively.
\item  $\theta_{\gamma\gamma}^*$ cut is varied by  $\pm3^\circ$, corresponding to $\sim 1\sigma$.
\item Missing mass cut (MM) is tested by varying the cut by $\pm2.0$
  MeV, $\sim 1\sigma$. For this cut a stronger dependence of the
  parameters on the cut was noted. This has been further investigated
  by performing the Dalitz plot parameter fit for one parameter at a
  time, for each step, and keeping the other parameters fixed at the
  value for the standard result. Since the dependence was reduced when
  varying just one parameter, we conclude that it is mostly due to
  the correlations between parameters.
\item Event classification procedure (ECL) is investigated by using a
  prescaled data sample without the event classification bias
  (collected with prescaling factor $1/20$). The fraction of events
  remaining in each Dalitz plot bin after the event classification
  conditions varies between 94\% and 80\% for different bins and it is
  very well described by the MC within the errors. The analysis of the
  prescaled data follows the standard chain.  The systematic error is
  extracted as half the difference between the results of the analysis
  with and without the event classification procedure.
\end{itemize}
Unless stated otherwise the systematic error is calculated as the difference between the two tests and the standard result. If both differences have the same sign, the asymmetric error is taken with one boundary set at zero and the other at the largest of the differences.
The resulting systematic error contributions for the Dalitz plot parameters for
the sets \#4 and \#3 are summarized in Tab.~\ref{tab:sys_summ_abdfg}
and Tab.~\ref{tab:sys_summ_abdf}, respectively. The results for the charge asymmetries are summarized in Tab.~\ref{tab:sys_asym}.

\begin{table}[htbp]
\renewcommand{\arraystretch}{1.3}
\caption{Summary of the systematic errors for $a,b,d,f,g$ parameters (fit $\#4$ ).  \label{tab:sys_summ_abdfg}}
\centering
\begin{tabular}{l | r r r r r }
\hline\hline
syst. error ($\times10^4$)& $\Delta a$ &  $\Delta b$& $\Delta d$ &$\Delta f$& $\Delta g$\\
\hline
{\bf EGmin} & $\pm 6$ & $\pm 12$ &  $\pm 10$ &  $\pm 5$  & $\pm 16$\\
{\bf BkgSub}  & $\pm 8$ &$ \pm 7 $&$\pm  11$&$\pm 6$&$\pm 38$\\
{\bf BIN}  & $\pm 17$ &$ \pm 13 $&$\pm  9$&$\pm 36$&$\pm 44$\\ 
$\theta_{+\gamma},\theta_{-\gamma}$ cut & $^{+0}_{-1}$ & $^{+0}_{-2}$ & $^{+2}_{-2}$ & $^{+3}_{-0}$ & $^{+3}_{-2}$ \\
$\Delta t_e$ cut& $_{-11}^{+\ 6}$ & $_{-\ 1}^{+12}$ & $_{-\ 1}^{+18}$ & $_{-8}^{+3}$ & $_{-54}^{+26}$\\
$\Delta t_e-\Delta t_\pi$ cut & $\pm 0$ & $_{-1}^{+0}$ & $_{-1}^{+3}$ & $\pm 0$ & $_{-1}^{+2}$ \\
$\theta_{\gamma\gamma}^*$ cut& $^{+14}_{-\ 5}$&$^{+2}_{-1}$&$^{+21}_{-12}$&$^{+\ 5}_{-25}$&$^{+26}_{-38}$\\
{\bf MM} & $^{+\ 8}_{-10}$&$^{+46}_{-43}$&$^{+49}_{-45}$&$^{+57}_{-62}$&$^{+100}_{-\ 92}$\\
{\bf ECL} & $\pm 0$ & $\pm 8$ & $\pm 6$ &$\pm9$& $\pm 12$ \\ \hline
{\bf TOTAL} & $^{+26}_{-25}$ &$ ^{+52}_{-48}$ & $^{+59}_{-50}$ &$^{+69}_{-77} $ & $^{+123}_{-129}$\\
\hline\hline
\end{tabular}
\end{table}
\renewcommand{\arraystretch}{1.0}
\begin{table}[htpb]
\renewcommand{\arraystretch}{1.3}
\caption{Summary of the systematic errors for $a,b,d,f$ parameters (fit $\#3$). \label{tab:sys_summ_abdf}}
\centering
\begin{tabular}{l | r r r r r }
\hline\hline
syst. error ($\times 10^{4}$)& $\Delta a$ &  $\Delta b$& $\Delta d$ &$\Delta f$\\
\hline
{\bf EGmin}  & $\pm 9$ & $\pm 10$ &  $\pm 6$ &  $\pm 0$ \\
{\bf BkgSub}   & $\pm 1$ &$ \pm 5 $&$\pm  6$&$\pm 8$\\ 
{\bf BIN}  & $\pm 9$ &$ \pm 14 $&$\pm  9$&$\pm 26$\\ 
$\theta_{+\gamma},\theta_{-\gamma}$ cut & $_{-1}^{+0}$ & $_{-2}^{+0}$ & $_{-1}^{+1}$ & $_{-0}^{+4}$ \\
$\Delta t_e$ cut& $_{-6}^{+0}$  & $_{-\ 6}^{+14}$ & $_{-0}^{+7}$ & $_{-15}^{+19}$\\
$\Delta t_e-\Delta t_\pi$ cut & $\pm 0$ & $_{-1}^{+0}$ & $_{-0}^{+3}$ & $\pm 0$ \\
$\theta_{\gamma\gamma}^*$ cut& $^{+6}_{-0}$&$^{+1}_{-1}$&$^{+14}_{-\ 8}$&$^{+\ 0}_{-13}$\\
{\bf MM} & $^{+10}_{-10}$&$^{+39}_{-36}$&$^{+31}_{-26}$&$^{+28}_{-35}$\\
{\bf ECL} & $\pm 2$ & $\pm 9$ & $\pm 9$ &$\pm13$\\ \hline
{\bf TOTAL} & $^{+18}_{-18}$ &$ ^{+46}_{-41}$ & $^{+38}_{-31}$ &$^{+45}_{-51} $\\
\hline\hline
\end{tabular}
\end{table}
\renewcommand{\arraystretch}{1.0}
\begin{table}[htbp]
\renewcommand{\arraystretch}{1.3}
\caption{Summary of the systematic errors for the asymmetries.  \label{tab:sys_asym}}
\centering
\begin{tabular}{l | r r r r r }
\hline\hline
syst. error ($\times10^5$)& $\Delta A_{LR}$ &  $\Delta A_Q$& $\Delta A_S$ \\
\hline
{\bf EGmin} & $\pm 1$ & $\pm 0$ &  $\pm 4$ \\
{\bf BkgSub}  & $\pm 5$ &$ \pm 3 $&$\pm  16$\\
$\theta_{+\gamma},\theta_{-\gamma}$ cut & $^{+2}_{-0}$ & $^{+0}_{-2}$ & $^{+2}_{-0}$ \\
 $\Delta t_e$ cut& $_{-92}^{+49}$ & $_{-22}^{+48}$ & $_{-15}^{+\ 7}$ \\
$\Delta t_e-\Delta t_\pi$ cut& $^{+0}_{-2}$ & $_{-0}^{+3}$ & $_{-1}^{+0}$  \\
$\theta_{\gamma\gamma}^*$ cut& $^{+\ 1}_{-57}$ & $^{+3}_{-4}$ & $^{+0}_{-8}$ \\
{\bf MM} & $^{+0}_{-4}$&$^{+0}_{-1}$&$^{+1}_{-2}$\\
{\bf ECL} & $\pm 9$ & $\pm 0$ & $\pm 25$ \\ \hline
{\bf TOTAL} & $^{+\ 50}_{-109}$ &$ ^{+48}_{-23}$ & $^{+31}_{-35}$\\
\hline\hline
\end{tabular}
\end{table}
\renewcommand{\arraystretch}{1.0}

\section{Discussion}

The final results for the Dalitz plot parameters, including systematic
effects, are therefore:
\begin{align*}
 a =-&1.095\pm0.003^{+0.003}_{-0.002}\\
b= +&0.145\pm0.003\pm 0.005\\
d=+&0.081\pm0.003^{+0.006}_{-0.005}\\
f=+&0.141\pm0.007^{+0.007}_{-0.008}\\
g=-&0.044\pm0.009^{+0.012}_{-0.013}\\
\end{align*}
including the $g$ parameter. With $g$ parameter set to zero the
results are:
\begin{align*}
 a &=-1.104\pm0.003\pm0.002\\
b&= +0.142\pm0.003^{+0.005}_{-0.004}\\
d&=+0.073\pm0.003^{+0.004}_{-0.003}\\
f&=+0.154\pm0.006^{+0.004}_{-0.005}.\\
\end{align*}

These results confirm the tension with the theoretical calculations on
the $b$ parameter, and also the need for the $f$ parameter.  In
comparison to the previous measurements shown in
Tab.~\ref{tab:prev_results}, the present results are the most
precise and the first including the $g$ parameter. The improvement
over KLOE(08) analysis comes from four times larger statistics and
improvement in the systematic uncertainties which are in some cases
reduced by factor $2-3$. 
The major improvement in the systematic
  uncertainties comes from the analysis of the effect of the Event classification  with
  an unbiased prescaled data sample.

The final values of the charge asymmetries are all consistent with
zero:
\begin{align*}
 A_{LR} &=(-5.0 \pm 4.5 ^{+5.0}_{-11}) \cdot 10^{-4}\\
A_Q&= (+1.8 \pm 4.5 ^{+4.8}_{-2.3})\cdot 10^{-4}\\
A_S&= (-0.4\pm 4.5 ^{+3.1}_{-3.5})\cdot 10^{-4}.\\
\end{align*}
The systematic and statistical uncertainties are of the same size
except for the $A_{LR}$ which is dominated by the systematic uncertainty 
due to the description of the Bhabha background.  

\begin{acknowledgments}


We warmly thank our former KLOE colleagues for the access to the data collected during the KLOE data taking campaign.
We thank the DA$\Phi$NE team for their efforts in maintaining low background running conditions and their collaboration during all data taking. We want to thank our technical staff: 
G.F. Fortugno and F. Sborzacchi for their dedication in ensuring efficient operation of the KLOE computing facilities; 
M. Anelli for his continuous attention to the gas system and detector safety; 
A. Balla, M. Gatta, G. Corradi and G. Papalino for electronics maintenance; 
M. Santoni, G. Paoluzzi and R. Rosellini for general detector support; 
C. Piscitelli for his help during major maintenance periods. 
This work was supported in part by the EU Integrated Infrastructure Initiative Hadron Physics Project under contract number RII3-CT- 2004-506078; by the European Commission under the 7th Framework Programme through the `Research Infrastructures' action of the `Capacities' Programme, Call: FP7-INFRASTRUCTURES-2008-1, Grant Agreement No. 227431; by the Polish National Science Centre through the Grants No.\
2011/03/N/ST2/02652,
2013/08/M/ST2/00323,
2013/11/B/ST2/04245,
2014/14/E/ST2/00262,
2014/12/S/ST2/00459.

\end{acknowledgments}
\appendix

\section{Acceptance corrected data\label{sec:APPA}}

With a smearing matrix close to diagonal and the smearing to and from
nearby bins symmetrical, the acceptance corrected data can be used
instead of dealing with the smearing matrix. This representation has
the advantage of being much easier to compare directly with
theoretical calculations. The acceptance corrected signal content in
each bin of the Dalitz plot is obtained by dividing the background
subtracted content, $N_i$, by the corresponding acceptance,
$\epsilon_i$.  The acceptance is obtained from the signal MC by
dividing the number of reconstructed events allocated to the bin $i$ by
the number of generated (unsmeared) signal events in that bin.

The fit to extract the Dalitz plot parameters values is done now by minimizing 
\begin{equation}
\chi^2 = \sum_{i=1}^{n} \left( \frac{N_i/\epsilon_i - N_{T,i}}{\sigma_i}    \right)^2 \label{eq:chi2_acccorr}
\end{equation}
where the sum includes only bins completely inside the Dalitz plot
boundaries and $ N_{T,i} = \int\int |A(X,Y)|^2 dX_i dY_i$.  The statistical uncertainty 
$\sigma_i$ includes contributions from the experimental data, the
background estimated from MC and the efficiency.  The fitted Dalitz plot
parameters
using the acceptance corrected data are presented in
Tab.~\ref{tab:dpfit_results} as sets $\#5$ with  $g$ parameter and
$\#6$ with $g = 0$.  The results are identical within statistical
uncertainties with the values obtained using the smearing
matrix. Therefore the acceptance corrected data can be used to
represent the measured Dalitz plot density if one neglects
systematical uncertainties.
The table containing Dalitz plot
acceptance corrected data (normalized to the content of the 
$X_c=0.0,Y_c= 0.05$ bin), is provided as a supplementary
material (file \texttt{DPhist\_acccorr.txt}). 
The correlation matrix for the fit \#5 reads:
\begin{center}
\begin{tabular}{ c |l l l l }
 & \multicolumn{1}{c}{$b$} &\multicolumn{1}{c}{$d$}&\multicolumn{1}{c}{$f$}&\multicolumn{1}{c}{$g$}\\
\hline
$a$ &  $-0.110$&  $+0.006$& $-0.849$& $-0.512$\\
$b$ &    & $+0.397$& $-0.216$& $-0.239$\\
$d$ &  &  &$-0.133$& $-0.537$\\
$f$ &  & &  & $+0.380$.\\
\end{tabular}
\end{center}

\bibliographystyle{./lisstyle}
\bibliography{./kloestuff,./eta3pi}

\end{document}